\documentclass[11pt,a4paper,amssymb,amsmath, tightenlines]{article}

\usepackage{amsmath, amssymb, amsthm, latexsym, mathrsfs, url, color, epsfig, graphics, float, setspace}
\usepackage{sectsty}
\usepackage[T1]{fontenc}
\usepackage[bitstream-charter]{mathdesign}
\usepackage{caption,subcaption}

\usepackage{geometry,enumerate}

\usepackage{graphicx}				
	\def\beq{\begin{equation}}
\def\eeq{\end{equation}}
\newcommand{\barr}{\begin{array}}
\newcommand{\earr}{\end{array}}
\newcommand{\bqr}{\begin{eqnarray}}
\newcommand{\bqq}{\begin{eqnarray*}}
\newcommand{\eqr}{\end{eqnarray}}
\newcommand{\eqq}{\end{eqnarray*}}
\newcommand{\rd}{\textrm{d}}
\newcommand{\an}{\textrm{and}}

\newcommand{\e}{\textrm{e}}

\newcommand{\oo}{\hat{\textrm{o}}}

\newcommand{\E}{\mathbb{E}}

\newcommand{\PR}{\mathbb{P}}
\newcommand{\RR}{\mathbb{R}}
\newcommand{\sign}{\textrm{sign}}

\numberwithin{equation}{section}

\theoremstyle{plain}
\newtheorem{thm}{Theorem}[section] 

\theoremstyle{definition}
\newtheorem{defn}[thm]{Definition} 

\newtheorem{prop}[thm]{Proposition}

\usepackage{doc}
\usepackage{url}
\usepackage{graphicx}
\usepackage{epstopdf}

\title{Dark-Pool Perspective of Optimal Market Making\footnotetext{ The authors thank  J. Walton for useful discussions at the beginning of this work and are grateful to C.A. Garcia Trillos for comments which improved this paper.}}
\author{M. Alessandra Crisafi$^{\ast}$ and Andrea Macrina$^{\ast\, \dagger}$ \\ 
$^{\ast}$Department of Mathematics, University College London\\
$^{\dagger}$Department of Actuarial Science, University of Cape Town }

\begin{document}

\maketitle
\begin{abstract}
\noindent We consider a finite-horizon market-making problem faced by a dark pool that executes  incoming buy and sell orders. The arrival flow of such orders is  assumed to be random and, for each transaction, the dark pool earns a per-share commission no greater than the half bid-ask spread. Throughout the entire period, the main concern is inventory risk, which increases as the number of held positions  becomes critically small or large. The dark pool can control its inventory by choosing the size of the commission for each transaction, so to encourage, e.g., buy orders instead of sell orders. Furthermore, it can submit lit-pool limit orders, of which execution is uncertain, and market orders, which are  expensive. In either case, the dark pool risks an information leakage, which we model via a fixed penalty for trading in the lit pool. We solve a double-obstacle impulse-control problem associated with the optimal management of the inventory, and we show that the value function is the unique viscosity solution of the associated system of quasi variational inequalities. We explore various  numerical examples of the proposed model, including one that admits a semi-explicit solution. \\ \vspace{-0.2cm}\\
{\bf Keywords:} Market-making, inventory risk, impulse-control problem, quasi variational inequality, viscosity solutions.

\vspace{.25cm}
\noindent
\end{abstract}
\section{Introduction}
Market makers are liquidity providers. They set bid and ask quotes and trade with impatient investors who seek to immediately buy or sell a certain quantity of a stock by market orders. The market maker is willing to hold a non-zero inventory while earning the spread from each transaction. Holding a non-zero inventory carries an intrinsic risk associated to the unpredictable changes to which the stock price is subject. This risk is further increased by a potential information asymmetry due to which the market maker trades in the wrong direction. 

In this paper we consider the optimal management of a dark pool inventory and the pool's role as a market maker. A dark pool is an alternative trading venue where participants do not release their identity and benefit of advantageous prices. While optimal-liquidation problems  involving dark pools have been treated in  recent works by, e.g., Kratz \& Sch\"{o}neborn (2013),  Horst \& Naujokat (2014), Crisafi \& Macrina (2014), and  Graewe et al. (2015), as far as we know dark pools  have not been studied in a market-making context elsewhere. In our situation the dark pool (i) executes incoming buy and sell orders posted by its clients and (ii) may post limit and market orders in a lit pool to control the level of its inventory.  This view can be justified by  the increasing competition between dark platforms and the reputational benefit deriving from the speed of execution and price advantage. Also, by executing large orders placed by its (institutional) clients, the dark pool acquires ``reserved'' information about the traded asset. Furthermore, the dark pool would rather avoid the situation where an investor (client) resorts to the lit pool and thus potentially moves the price against the dark pool's holding. We formulate a double-obstacle impulse-control problem and we use the viscosity theory to characterise the solution of the associated system of quasi variational inequalities (QVIs). We refer the reader to Crandall et al. (1992), Fleming \& Soner (2006) and Pham (2009) for a comprehensive treatment of viscosity solutions.

Previous work on market making includes  Amihud \& Mendelson (1980) who, based on Garman (1976), relate the  bid-ask prices to the share holding of a risk-neutral agent. They find a relationship between the optimal quotes and the distance from the ``preferred'' inventory position. Stoll (1978) considers a two-period model in which a risk-averse agent supplies liquidity and maximises his expected utility.  Ho \& Stoll (1981) use the dynamic programming principle (DPP) to obtain the optimal quotes which maximise the terminal wealth in a single-dealer market. The recent evolution in financial markets, arisen with algorithmic and high-frequency trading, has shifted the optimal market-making problem to an order-driven market environment where optimal quotes and trading strategies are computed and submitted by electronic machines. For example Avellaneda \& Stoikov  (2008)  adapt the market-making framework by Ho \& Stoll (1981)  to a limit order book (LOB). They consider the maximisation of the agent's expected terminal wealth and consider both, the finite and the infinite-time cases. They model the arrival of buy and sell orders by Poisson processes and the dynamics of the mid-price by an arithmetic Brownian motion. They find the HJB PDE by means of the DPP and propose an approximation of the optimal quotes via asymptotic expansions. This type of problem has been investigated elsewhere, too. The works by Cartea \& Jaimungal (2012) on risk metrics and by Cartea et al. (2013, 2014) consider ambiguity and self-exciting processes, respectively.  Gu\'eant et al. (2013) deal with the inventory risk and reduce a complex optimisation problem to a system of ODEs. Guilbaud \& Pham (2013) consider a market maker who continuously submits limit orders at the best quoted prices and resorts to market orders when the inventory becomes too large. They numerically solve a finite-time impulse-control problem and find the optimal order sizes and quotes to be posted in the lit pool.

The paper is organised as follows. After the introduction, in Section \ref{eq:model}, we present the inventory and cash processes of the dark pool and we describe its trading strategies. In Section \ref{eq:optim}, the optimisation problem is introduced and, by making use of DPP, we derive the HJB equation. Section \ref{eq:num} is devoted to a numerical analysis of the dark pool strategies. All propositions, necessary to characterise the value function by the unique viscosity solution of the system of QVIs, and their proofs can be found in the appendix.
\vspace{-0.2cm}
\section{Dark pool as a market maker\label{eq:model}}
We consider a dark pool that executes incoming buy and sell orders by its clients over a finite period of time $0\leq t\leq T<\infty$ and that may resort to the lit pool if its inventory becomes critically small or large. As a reward for the service provided,  the dark pool chooses a per-share commission $\delta^{a}$ for sell orders and $\delta^{b}$ for buy orders, which shall not exceed the lit-pool half spread. The dark-pool order flow is affected by changing $\delta^{a}$ and $\delta^{b}$. We may assume for example that the dark pool has a positive inventory at time $u\in[t,T]$. It can make buy orders more attractive than sell orders, so to rebalance its position on the stock. In particular, by setting $\delta^{b}$ close to the mid-price and $\delta^{a}$ far away from the mid-price (close to the best bid price), it encourages buy orders while sell orders are discouraged.
At each time $u\in[t,T]$, we thus consider three  options for the inventory management: i) the dark pool's order flow may be controlled by optimally choosing the per-share commission size, ii) a limit order---of which execution is uncertain---is posted in the lit pool , or iii) a costly market order is submitted to the lit pool.  Our goal is to obtain the critical  levels of the inventory for which  market orders or limit orders are optimal.

We fix a filtered probability space $(\Omega,\mathcal F,\{\mathcal F_u\}_{0\leq u\leq T},\mathbb{P})$ satisfying the usual conditions and augmented by all $\PR$-null sets. We model the LOB bid-ask half spread by a continuous-time Markov chain  $\{k(u)\}_{t\leq u\leq T}$ with discrete state space $\mathbb K:=\{0,\delta,2\delta\dots n\delta\}$. The chain is generated by $\{Q\}=(r_{ij})$ such that $\PR[k(u+\rd u)=j|k(u)=i]=  r_{ij}\rd u$ and $\PR[k(u+\rd u)=i|k(u)=i]= 1+ r_{ii}\rd u$, with $r_{ij}\geq 0$ for all $j\neq i$ and $r_{ii}=-\sum_{j\neq i}r_{ij}$. The LOB mid-price is  defined by
\begin{equation}
S(t_1)=s+\int_{t}^{t_1}\mu\left(u,S(u)\right)\rd u+\int_{t}^{t_1}\sigma\left(u,S(u)\right)\rd W(u),
\label{eq:midprice}
\end{equation}
where $\{W(u)\}_{t\leq u\leq T}$ is a standard $\{\mathcal F_u\}$-Brownian motion. At  $t_{1}\in[t,T]$, the best bid and ask prices are given  by $S^{b}(t_{1})=S(t_{1})-k(t_{1})$ and $S^{a}(t_{1})=S(t_{1})+k(t_{1})$, respectively.  

We let $a$ and  $b$ be $\RR_+\cup\{0\}$-valued i.i.d. random variables with finite second moment. These variables model the size of incoming sell and buy orders in the dark pool. We define the inventory process $\{X(u)\}_{t\leq u\leq T}$  by
\begin{equation}
 X(t_1)=x+\int_t^{t_1}f\left(u,X(u),a\right)\,\rd N^a(u)-\int_t^{t_1}f\left(u,X(u),b\right)\,\rd N^b(u),
\label{eq:inventory}
\end{equation}
where $\{N^a(u)\}$ and $\{N^b(u)\}$ are independent Poisson processes with intensities $\lambda^{a}_\delta=\lambda^a(\delta^{a})$ and $\lambda^{b}_\delta=\lambda^b(\delta^{b})$, respectively. We allow for short-selling and thus, at any time $u\in[t,T]$, we have $\sign[X(u)]=\{-1,0,1\}$.  Furthermore, we model the cash process $\{Y(u)\}_{t\leq u\leq T}$  by
\begin{eqnarray}
Y(t_1)=y&+&\int_t^{t_1}\!\!\!\!f_1\big(u,X(u),Y(u),S(u),\delta^{b}(u),b\big)\rd N^b(u)\nonumber\\
&-&\int_t^{t_1}\!\!\!\!f_1\big(u,X(u),Y(u),S(u),\delta^{a}(u ),a\big)\rd N^a(u).
\label{eq:cash}
\end{eqnarray}
Equations ($\ref{eq:inventory}$) and ($\ref{eq:cash}$) are strongly connected. For example, an incoming buy order of size $b$ reduces the inventory by $f\left(u,X(u),b\right)$ and increases the cash amount by $f_1\big(u,X(u),Y(u),S(u),\delta^{b}\!,b\big)$. The analogous holds for incoming sell orders. For technical reasons, we consider a bounded domain for both, the inventory and the cash processes, by letting  $(x,y)\in\mathcal O\subset\RR^{2}$.We let $\mathcal O:=[\underline{X}\,,\overline{X}]\times[\underline{Y}\,,\overline{Y}]$,  where $\underline{X}$\,, $\overline{X}$, $\underline{Y}$\,, and $\overline{Y}$ are real-valued constants. Such an assumption  is supported by the following financial interpretation:  the market maker, i.e. the dark pool,  is subject to regulations constraints (e.g. internal risk-management) which make it hard to hold or short-sell an amount of shares bigger than a fixed authorised quantity. Also, the market maker  exits the market  whenever the cash process has reached  either the lower bound (bankruptcy) or  an upper bound (gains target).  The dark pool can resort to the lit pool to liquidate (respectively refill) part of the inventory; we assume that it can not post speculative orders. This means that at time $u\in[t,T]$ a buy order can be posted  only if $X(u)<0$ while a sell order can be posted if $X(u)>0$.
\vspace{-0.35cm}
\subsubsection*{Limit orders strategy}
\vspace{-0.15cm}
The dark pool (market maker) can post a limit order by specifying a quantity $\eta$ and a limit price $S\pm(k+\kappa)$, where $\kappa$ is the optimal distance from the best price, at which it wants to buy or sell. We only consider immediate-or-cancel orders and we model their execution percentage by a $[0,1]$-valued random variable $z$. In particular, if a limit order  is posted at time $\tau_j$, for $j=1,2\dots$, then it impacts the inventory and the cash processes as follows:
\begin{equation}
\begin{split}
&X\big(\tau_j\big)=X\big(\tau_{j^{-}}\big)+\Gamma\big(\eta_j,X\big(\tau_j\big),z\big),\\
&Y\big(\tau_j\big)=Y\big(\tau_{j^{-}}\big)+\chi\big(\eta_j,X\big(\tau_j\big),z,S(\tau_j),k(\tau_{j}),\kappa_{j}\big).
\end{split}
\end{equation}
For a limit buy order, the function $\Gamma$ is non-negative valued and the function $\chi$ is non-positive valued. In fact, a limit buy order, if executed, increases the inventory and reduces the cash amount. The contrary holds for limit sell orders. We will state these assumptions rigorously in the appendix. We let $\mathcal T_T$ be the set of stopping times not greater than T, and $\mathcal N$ be the set of all admissible control actions. A limit-order  strategy is a collection of stopping times and actions $L=(\tau_j,\eta_j,\kappa_{j})\in\mathcal T_T\times\mathcal N\times[0,\bar\kappa]$.
\vspace{-0.35cm}
\subsubsection*{Market orders strategy}
\vspace{-0.15cm}
Alternatively, the market maker can submit  a market order, which (i) is  more expensive and (ii) benefits of sure execution as it is matched with existing limit orders. A market order of size $\xi_i$ posted at a time $\rho_i$ impacts the inventory and cash processes as follows
\begin{equation}
\begin{split}
&X\big(\rho_i\big)=X\big(\rho_{i^{-}}\big)+\Lambda\big(\xi_i,X\big(\rho_i\big)\big),\\
&Y\big(\rho_i\big)=Y\big(\rho_{i^{-}}\big)+c\big(\xi_i,X\big(\rho_i\big),S(\rho_i),k(\rho_{i})\big).
\end{split}
\end{equation}
For a market buy order, the function $\Lambda$ is non-negative and the function $c$ is non-positive. The contrary holds for market sell orders. A market-order  strategy is a collection of stopping times and actions $M=(\rho_i,\xi_i)\in\mathcal T_T\times\mathcal X$.
\vspace{-0.25cm}
\section{Optimisation problem and viscosity solution \label{eq:optim}}
We consider the problem of maximising expected terminal cash subject to total liquidation (the problem is the same for a pre-specified non-zero terminal inventory) of the remaining inventory  via a market order. In defining the objective function, we are led by Guilbaud \& Pham (2013). We let the stopping time $\tau^{*}$ be the first time the state variables exit  the set $\mathcal O$, such that
$$
\tau^{*}=\inf\,\{u>t\,|\,(X(u),Y(u))\notin\mathcal O\}\wedge T.
$$
We define the value function by
\begin{equation}
\begin{split}
V\left(t,x,y,s;k\right)=\sup_{D,L,M}\E\Biggl[&\int_{t}^{\tau^{*}}g(u,X(u))\,\rd u-\sum_{t<\rho_i\leq T}\epsilon_m-\sum_{t<\tau_j\leq T}\epsilon_l\\
&+U\left(X(\tau^{*}),Y(\tau^{*}),S(\tau^{*}),k(\tau^{*})\right)\Biggr],
\end{split}
\label{eq:valuefunction}
\end{equation}
where $D=(\delta^{a},\delta^{b})$, the function $U$ is the time-$\tau^{*}$ liquidation revenues and $g$ is a running penalty for the risk of holding the inventory. In particular, (i) if $g$ is negative-valued, the dark pool is risk-averse, (ii) if $g$ is equal to zero, then it is risk-neutral, and (iii) if $g$ is positive, then the dark pool is risk-prone. In the summations we include the penalties for trading in the lit pool, where $\epsilon_m>\epsilon_l>1$. Throughout the paper  we write, whenever possible, the vector of state variables ${\boldsymbol x}:=[x,y,s]\in\mathcal O\times\RR_{+}=:\mathcal S$ and we let $\mathcal T_{\tau^*}$ be the set of all stopping times less than $\tau^*$. Equation (\ref{eq:valuefunction}) satisfies the DPP, see Fleming \& Soner (2006). That is, for all $\tau\in\mathcal T_{\tau^*}$,
\begin{equation}
V\left(t,{\boldsymbol x};k\right)=\sup_{D,L,M}\E\left[\int_{t}^\tau g(u,X(u))\rd u-\sum_{t<\rho_i\leq \tau}\epsilon_m-\sum_{t<\tau_j\leq \tau}\epsilon_l+V\left(\tau,{\boldsymbol X}_{t,\,{\boldsymbol x}}(\tau);k(\tau)\right)\right].
\label{eq:dpp}
\end{equation}
This is an optimal double-obstacles impulse control problem. We define the non-local operators $\mathcal L$ and $\mathcal M$, for limit and market orders respectively, by
$$
\mathcal L V\left(t,{\boldsymbol x};k\right)=\sup_{\eta\in\mathcal N,\kappa\in[0,\bar\kappa]}\int_0^1V\left(t,x+\Gamma(\eta,x,z),y+\chi(\eta,x,z,s,k,\kappa),s;k\right)\ell^{(\kappa)}_{z}(z)\rd z-\epsilon_l,
$$
and 
$$
\mathcal M V\left(t,{\boldsymbol x};k\right)=\sup_{\xi\in\mathcal X}V\left(t,x+\Lambda(\xi,x),y+c(\xi,x,s,k),s;k\right)-\epsilon_m.
$$
We define the operator $\mathcal A \left(t,{\boldsymbol x},k,p,q,M,\phi\right)=\sup_{D}\bar{\mathcal A} \left(t,{\boldsymbol x},k,p,q,M,\phi,D\right)$ by
\begin{equation*}
\bar{\mathcal A} \left(t,{\boldsymbol x},k,p,q,M,\phi\right)=H(t,{\boldsymbol x},p,q,M)+\mathcal B^{k}_{a}(t,{\boldsymbol x},\phi)+\mathcal B^{k}_{b}(t,{\boldsymbol x},\phi)+\mathcal Q\phi(t,{\boldsymbol x};k),
\end{equation*}
\vspace{-0.3cm}
where
\begin{equation*}
\begin{split}
&\mathcal H(t,{\boldsymbol x},p,q,M):=p+\mu(t,s)q+\tfrac{1}{2}\sigma^{2}(t,s)M,\\
&\mathcal B^{k}_{a}(t,{\boldsymbol x},\phi) :=\!\!\!\!\sup_{\delta^{a}\in[0,k]}\!\!\!\!\mathcal  \lambda^{a}_\delta\!\!\int_{0}^\infty\!\!\!\!\!\!\!\left(\phi\left(t,x+f(t,x,a),y-f_{1}(t,{\boldsymbol x},\delta^{a},a),s;k\right)-\phi\left(t,{\boldsymbol x};k\right)\right)\ell_{a}(a)\rd a,\\
&\mathcal B^{k}_{b}(t,{\boldsymbol x},\phi):=\!\!\!\!\sup_{\delta^{b}\in[0,k]}\!\!\!\!\mathcal \lambda^{b}_\delta\!\!\int_{0}^\infty\!\!\!\!\!\!\!\left(\phi\left(t,x-f(t,x,b),y+f_{1}(t,{\boldsymbol x},\delta^{b},b),s;k\right)-\phi\left(t,{\boldsymbol x};k\right)\right)\ell_{b}(b)\rd b,\\
&\mathcal Q\phi(t,{\boldsymbol x};k)=\sum_{k'\neq k}r_{kk'}(\phi(t,{\boldsymbol x};k')-\phi(t,{\boldsymbol x};k)).
\end{split}
\end{equation*}
The value function $V(t,{\boldsymbol x};k)$ satisfies the QVI system
\begin{equation}
\min\left\{-g(t,x)-\mathcal A\left(t,{\boldsymbol x},k,\partial_{t }V,\partial_{s}V,\partial_{ss}V,V\right),(V-\mathcal MV)\left(t,{\boldsymbol x};k\right),(V-\mathcal L V)\left(t,{\boldsymbol x};k\right)\right\}=0,
\label{eq:qvi}
\end{equation}
on $[0,T)\times\mathcal S\times\mathbb K$. Equation (\ref{eq:qvi}) can be interpreted as follows: if $V-\mathcal MV>0$ and $V-\mathcal LV>0$, then  the value function can not be improved by  an impulse and thus no orders are submitted to the lit pool. As soon as $V-\mathcal MV<0$ or $V-\mathcal LV<0$, the value function is set to $V-\mathcal MV=0$ or $V-\mathcal LV=0$ and an impulse takes place. We thus consider intervention times ($\tau_j$ and $\rho_i$) and impulses ($\eta_j$ and $\xi_i$) by which the dark pool can control the evolution of the state variables $X(u)$ and $Y(u)$. For this purpose, we define the continuation region ($CR$), the limit orders impulse region ($LI$) and the market orders impulse region ($MI$) by
\begin{equation}
\begin{split}
CR&:=\left\{ \left(u,{\boldsymbol x},k\right)\in[0,T)\times \mathcal S\times\mathbb K:V>\mathcal L V\ \an\  V>\mathcal M V\right\},\\
LI&:=\left\{ \left(u,{\boldsymbol x},k\right)\in[0,T)\times \mathcal S\times\mathbb K:\mathcal L V=V\ \an\  \mathcal LV>\mathcal M V\right\},\\
MI&:=\left\{ \left(u,{\boldsymbol x},k\right)\in[0,T)\times \mathcal S\times\mathbb K:\mathcal M V=V\ \an\  \mathcal MV>\mathcal L V\right\}.
\label{eq:regions}
\end{split}
\end{equation}
For each $k\in\mathbb K$, let us consider the upper and lower semi-continuous envelopes of the function $V(\cdot;k)$ defined by
$$
V^{*}(t,{\boldsymbol{x}};k)=\lim_{t'\rightarrow t}\sup_{{\boldsymbol x}'\rightarrow {\boldsymbol x}}V(t',{\boldsymbol{x}}';k),\quad V_{*}(t,{\boldsymbol{x}};k)=\lim_{t'\rightarrow t}\inf_{{\boldsymbol x}'\rightarrow {\boldsymbol x}}V(t',{\boldsymbol{x}}';k).
$$
\begin{defn}
{\itshape A system of  functions $V:[0,T)\times\mathcal S\times\mathbb K\rightarrow\RR$ is a viscosity subsolution, (resp. supersolution), of (\ref{eq:qvi}) if  
\begin{equation*}
\begin{split}
\min\Big\{\!\!-g(\bar t,\bar x)\!- \mathcal A\left(\bar t,\bar{\boldsymbol x},\bar k,\partial_{t}\phi,\partial_{s}\phi,\partial_{ss}\phi,\phi\right)      ,&(V^{*}\!-\!\mathcal MV^{*})\left(\bar{t},\bar{\boldsymbol x};\bar k\right),\\
&(V^{*}\!-\!\mathcal L V^{*})\left(\bar{t},\bar{\boldsymbol x};\bar k\right)\!\!\Big\}\leq0,\\
\end{split}
\end{equation*}
\begin{equation*}
\begin{split}
\bigg(resp.\ \ \ \ \  \min\Big\{\!\!-g(\bar t,\bar x)\!- \mathcal A\left(\bar t,\bar{\boldsymbol x},\bar k,\partial_{t}\phi,\partial_{s}\phi,\partial_{ss}\phi,\phi\right)      ,&(V_{*}\!-\!\mathcal MV_{*})\left(\bar{t},\bar{\boldsymbol x};\bar k\right),\\
&(V_{*}\!-\!\mathcal L V_{*})\left(\bar{t},\bar{\boldsymbol x};\bar k\right)\!\!\Big\}\geq0\bigg),
\end{split}
\end{equation*}
where $\phi\in\mathcal C^{1,2,0}([0,T)\times\mathcal S\times\mathbb K)$ is such that $V^{*}(t,{\boldsymbol x},k)-\phi(t,{\boldsymbol x};k)$ (resp. $V_{*}(t,{\boldsymbol x},k)-\phi(t,{\boldsymbol x};k)$) attains its maximum (resp. minimum) at $(\bar{t},\bar{{\boldsymbol x}},\bar k)\subset[t,T)\times\mathcal S\times\mathbb K$.
}
\label{eq:def1}
\end{defn}
\begin{defn}
{\itshape A system of functions $V:[0,T)\times\mathcal S\times \mathbb K \rightarrow\RR$ is a viscosity subsolution (resp. supersolution) of (\ref{eq:qvi})  if 
\begin{equation*}
\begin{split}
&\min\Big\{\!\!-g(\bar t,\bar{{\boldsymbol x}})\!-\mathcal A(\bar t,\bar{\boldsymbol x},\bar k,p,q,M,V^{*}),(V^{*}\!-\!\mathcal MV^{*})(\bar t,\bar{{\boldsymbol x}};\bar k),(V^{*}\!-\!\mathcal L V^{*})(\bar t,\bar{{\boldsymbol x}};\bar k)\Big\}\leq0,\\
& \big(resp.\min\Big\{\!\!-g(\bar t,\bar{{\boldsymbol x}})\!-\mathcal A(\bar t,\bar{\boldsymbol x},\bar k,p,q,M,V_{*}),(V_{*}\!-\!\mathcal MV_{*})(\bar t,\bar{{\boldsymbol x}};\bar k),(V_{*}\!-\!\mathcal L V_{*})(\bar t,\bar{{\boldsymbol x}};\bar k)\Big\}\geq0\big),
\end{split}
\end{equation*}
where
\begin{enumerate}[(i)]
\item the function $\phi\in\mathcal C^{1,2,0}([0,T)\times\mathcal S\times\mathbb K)$ is such that $V^{*}(t,{\boldsymbol x};k)-\phi(t,{\boldsymbol x};k)$ (resp. $V_{*}(t,{\boldsymbol x};k)-\phi(t,{\boldsymbol x};k)$) attains its maximum (resp. minimum) at $(\bar{t},\bar{{\boldsymbol x}},\bar k)\subset[0,T)\times\mathcal S\times \mathbb K$,\\
\item  $(p,q,M)\in{\mathcal{\bar{P}}}^{2,+}V^{*}(\bar t,\bar{{\boldsymbol x}};\bar k)$ (resp. ${\mathcal{\bar{P}}}^{2,-}V_{*}(\bar t,\bar{{\boldsymbol x}};\bar k)$), where the set ${\mathcal{\bar{P}}}^{2,+}V$ is the closure of the second-order parabolic superjet ${\mathcal{{P}}}^{2,+}V$ defined by
\begin{equation*}
\begin{split}
{\mathcal{{P}}}^{2,+}V(\bar t,\bar{{\boldsymbol x}};\bar k):=\bigg\{\left(p,q,M\right)\!\!\in\RR^{3}:V\left(t,{\boldsymbol x};\bar k\right)\leq&\ V(\bar t,\bar{{\boldsymbol x}};\bar k)+p(t-\bar t)\\
&+q(s-\bar s)+\tfrac{1}{2}M(s-\bar s)^{2}\bigg\},
\end{split}
\end{equation*}
\item the set $\mathcal{P}^{2,-}V=-\mathcal{P}^{2,+}(-V)$ is the second-order parabolic subjet.
\end{enumerate}
\label{eq:def2}}
\end{defn}
\noindent We use Definition \ref{eq:def1} to prove the existence of a viscosity solution and Definition \ref{eq:def2} to prove the strong comparison result, which implies uniqueness, for discontinuous viscosity solutions. The results and the standing assumptions are found in the appendix. We note further that, since the control space is bounded, the terminal condition is regular and equal to $V(T,{\boldsymbol x};k)=U({\boldsymbol x},k)$. For further details, see e.g. Pham (2009).
\section{Numerical results\label{eq:num}}
\begin{figure}[H]
\hspace{-2cm}\includegraphics[width=500pt,height=200pt]{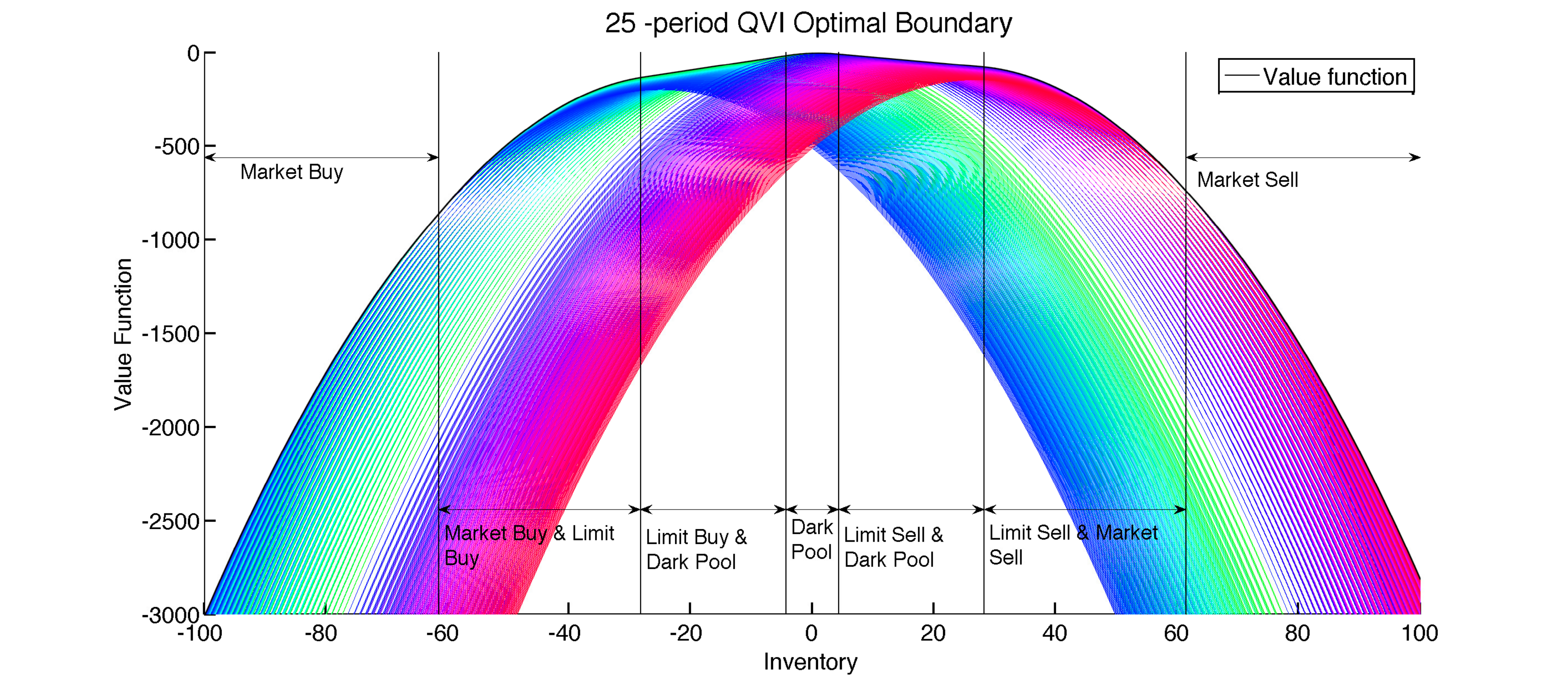}
\caption{\footnotesize{$n(n+1)(n+2)(n+3)/24=20,475$ distinguishable combinations of market, limit and dark pool orders. We set $\delta^{a}=\delta^{b}=0.25$, $\lambda^{a}=\lambda^{b}=0.5$, $k=0.5$, $\epsilon_{m}=5$, $\epsilon_{l}=3$, $p=0.92$.
}}
\label{eq:vf1}
\end{figure}

\begin{figure}[H]
\hspace{-0.8cm}\subcaptionbox{\footnotesize{$\epsilon_{l}<p(\epsilon_{m}+3k+pk)$}\label{fig:vf2}}{\includegraphics[width=235pt,height=150pt]{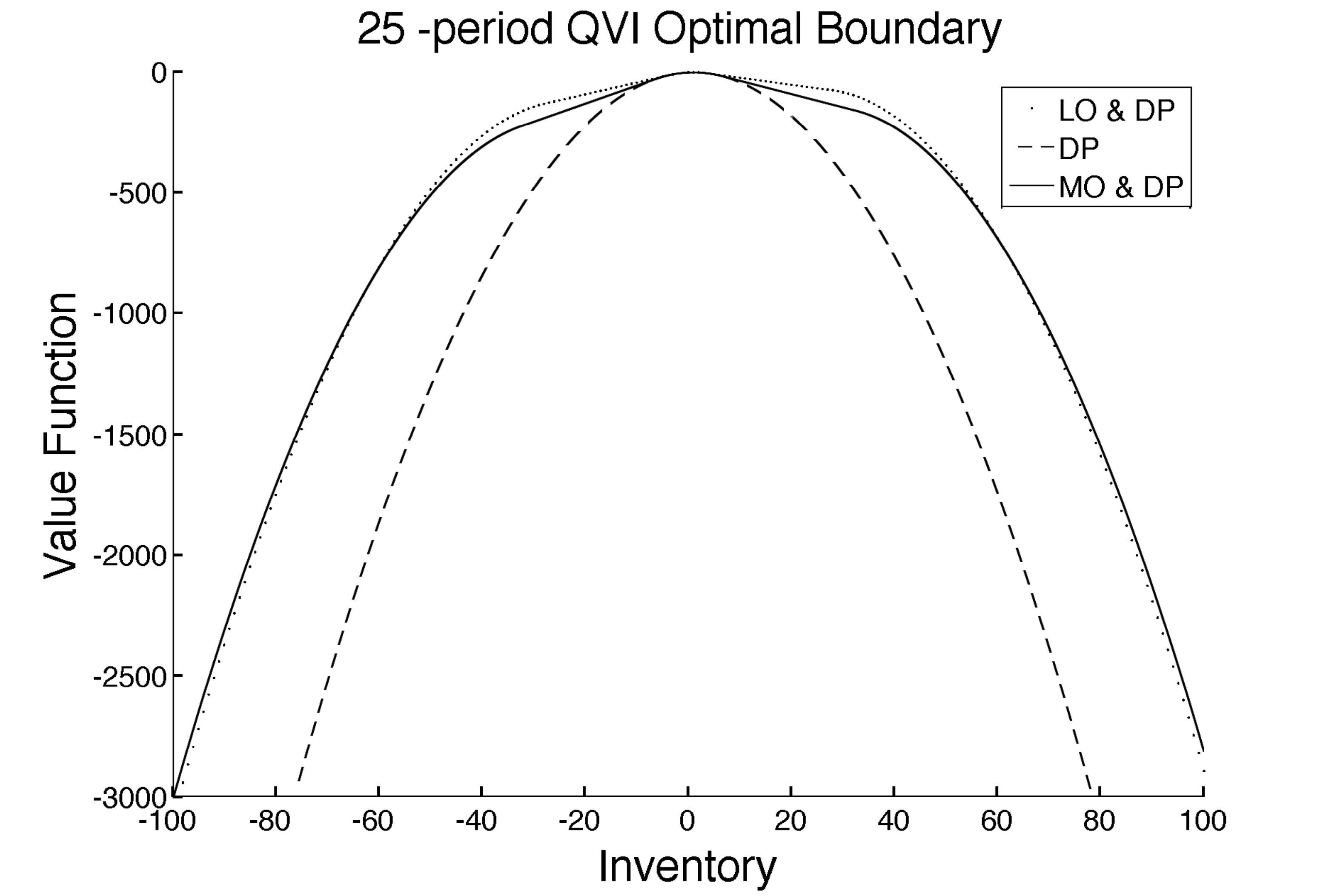}}
\hspace{-0.5cm}\subcaptionbox{\footnotesize{$\epsilon_{l}>p(\epsilon_{m}+3k+pk)$}\label{fig:vf3}}{\includegraphics[width=235pt,height=150pt]{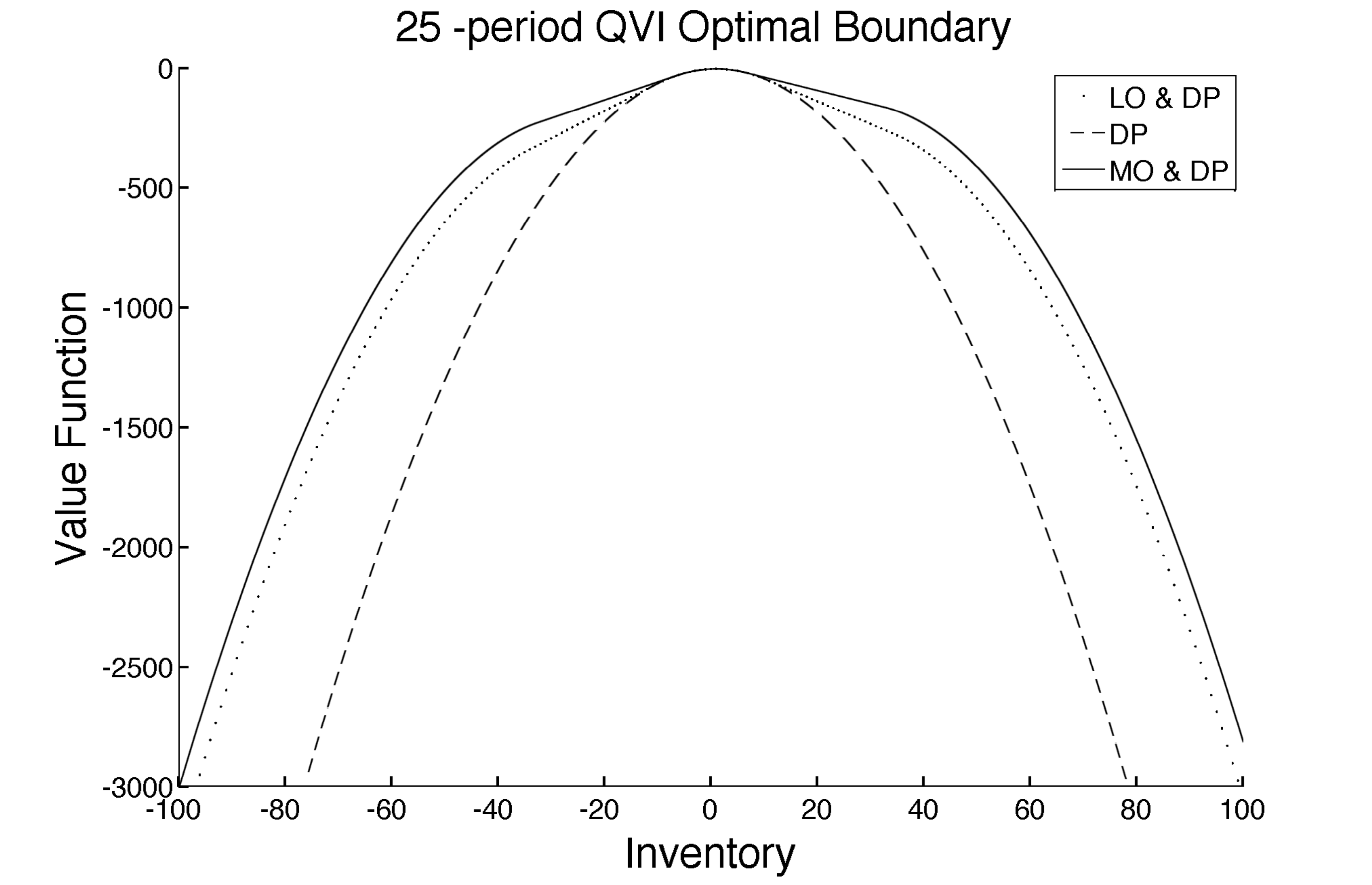}}
\caption{\footnotesize{Value function for dark pool only (dashed line), for limit orders and dark pool (dotted line), for market orders and dark pool (solid line). Limit orders perform better than market orders if $\epsilon_{l}<p(\epsilon_{m}+3k+pk)$}.}
\label{eq:vf23}
\end{figure}

In this section we explore some explicit examples of the impulse-control problem presented in this work and we analyse in details their qualitative features. We start by presenting a toy model, for which there exists a semi-explicit solution. Then we move forward by progressively removing the simplifying assumptions listed next. For the time being, we  suppose that
\begin{enumerate}[(a)]
\item the mid-price process follows an arithmetic Brownian motion,
\item dark-pool orders are unit-sized,
\item lit-pool orders are unit-sized,
\item the dark-pool commissions $\delta^{a}$ and $\delta^{b}$ are fixed  {\itshape a priori},
\item limit orders can not be partially executed (i.e. $z\in\{0,1\}$),
\item limit orders can only be posted at the best bid-ask prices (i.e. $\kappa=0$),
\item the dark pool observes the system and can intervene at fixed discrete times,
\item $g(u,x)=0$ and $U(x,y,s)=y+sx-kx^{2}$,
\item the bid-ask half spread is constant and set to $k$.
\end{enumerate}
In assumption (h), $g(u,x)=0$ means that the dark pool is risk-neutral. Moreover, given the choice of the function $U$, it seeks to maximise the terminal cash amount, subject to full liquidation of its inventory via a market order. Let us fix $t=t_{0}<t_{1}<t_{2} <\dots<t_{N}=T$ to be an equally-spaced partition of the time interval $T-t$, such that $t_{i+1}-t_{i}=\Delta>0$ for $i=0,1,\dots ,N-1$. We define the observed inventory and  cash processes by
\begin{eqnarray*}
X(t_{i+1})\!\!\!&=&\!\!\!X(t_{i})+(N^{b}(t_{i+1})-N^{b}(t_{i}))-(N^{a}(t_{i+1})-N^{a}(t_{i})),\\
Y(t_{i+1})\!\!\!&=&\!\!\!Y(t_{i})-(S(t_{i+1})-\delta^{b})(N^{b}(t_{i+1})-N^{b}(t_{i}))+(S(t_{i+1})+\delta^{a})(N^{a}(t_{i+1})-N^{a}(t_{i})).
\end{eqnarray*}
In the case that trades are executed only within the dark pool, the value function of the thus uncontrolled problem simplifies to
\begin{equation*}
V(t,x,y,s)=\E\left[Y(T)+X(T)S(T)- kX^{2}(T)\right]=y+xs+c_{1}(T-t)- k(x+c_{2}(T-t))^{2},
\end{equation*}
where
$
c_{1}=\lambda^{a}\delta^{a}+\lambda^{b}\delta^{b}- k\big(\lambda^{a}+\lambda^{b}\big)$
and
$c_{2}=\big(\lambda^{b}-\lambda^{a}\big)$.
Next, we introduce the possibility of submitting  unit-sized market and limit orders in the lit pool. At each time $t_{i}$, the dark pool checks whether it is more convenient to (i) execute trades in the dark pool only, (ii) submit a market order such that
$$
X(t_{i})=X(t_{i^-})+\xi(t_{i}),
$$
$$
Y(t_{i})=Y(t_{i^-})-\xi(t_{i})\left(S(t_{i})+\xi(t_{i}) k\right),
$$
where $\xi(t_{i})=\{+1,-1\}$ and $\epsilon_{m}>0$, or (iii) submit a limit order such that
$$
X(t_{i})=X(t_{i^-})+\eta(t_{i})z,
$$
$$
Y(t_{i})=Y(t_{i^-})-\eta(t_{i})\left(S(t_{i})-k\eta(t_{i})\right)z,
$$
where $\eta(t_i)=\{+1,-1\}$ and $z$ is a $\{0,1\}$-valued random variable with $\PR[z=1]=\ell(z_1)$ and $\PR[z=0]=\ell(z_0)$. Within this particular example,  a different time-ordering of the same actions does not influence the value function. We thus provide the following definition.
\begin{defn}
{\itshape Let $n\leq N$ and let ${\boldsymbol Q}:=\{Q_{m}^{a},Q_{m}^{b},Q_{l}^{a},Q_{l}^{b}\}\in \mathbb N^{4}$ be respectively the number of market sell (MS), market buy (MB), limit sell (LS) and limit buy (LB) orders  submitted in $[T-(n-1)\Delta,T]$, such that $Q_{m}^{a}+Q_{m}^{b}+Q_{l}^{a}+Q_{l}^{b}\leq n-1$. We say that two strategies ${\boldsymbol Q}_{1}$ and ${\boldsymbol Q}_{2}$ are {\bf \itshape distinguishable} if $\exists$ $Q_{1}\in {\boldsymbol Q}_{1}$ and $Q_{2}\in {\boldsymbol Q}_{2}$ such that $Q_{1}\neq Q_{2}$.}
\label{eq:ind}
\end{defn}
\noindent We have $n^{*}=\frac{1}{4!}\prod_{i=0}^{3}(n+i)$ distinguishable strategies for the time-interval  $[T-(n-1)\Delta,T]$. The situation can be summarised as follows: we have four objects and $n-1$ slots. We need to count how many combinations we may have, including repetitions, provided that the order does not count and some slots may stay empty. Such number $n^{*}$ is readily obtained:
$$
n^{*}=\sum_{i=0}^{(n-1)}\sum_{j=0}^{(n-1-i)}\sum_{k=0}^{(n-1-i-j)}\sum_{h=0}^{(n-1-i-j-k)}1=\frac{1}{4!}\prod_{i=0}^{3}(n+i).
$$
By the DPP, we can solve the above problem by means of backward recursion. In particular, let  ${\boldsymbol Q}$ be the { \itshape optimal} strategy for the interval $[T-(n-1)\Delta,T]$. Then the value function is specified by
\begin{equation*}
\begin{split}
V\big(T-(n-1)\Delta,x,y,s,Q_{m}^{a},Q_{m}^{b},Q_{l}^{a},Q_{l}^{b}\big) &= \,y+xs+c_{1}(T-(n-1)\Delta)-\big( k+\epsilon_{m}\big)\big(Q_{m}^{a}+Q_{m}^{b}\big)\\
&+\big(kp-\epsilon_{l}\big)\big(Q_{l}^{a}+Q_{l}^{b}\big)-k\big(x+\big(Q_{l}^{b}-Q_{l}^{a}\big)p+Q_{m}^{b}\\
&-Q_{m}^{a}+c_{2}(T-(n-1)\Delta)\big)^{2},
\end{split}
\end{equation*}
\\
where $\mathbb P[z=1]=p$ and $\mathbb P[z=0]=1-p$. The value function $V\big(T-n\Delta,x,y,s,Q_{m}^{a},Q_{m}^{b},Q_{l}^{a},Q_{l}^{b}\big)=V\big(T-n\Delta,Q_{m}^{a},Q_{m}^{b},Q_{l}^{a},Q_{l}^{b}\big )$ is given by
\\
\\
$$
\left\{ 
\begin{alignedat}{4}
&MB\!&&:&&V\big(Q_{m}^{a},Q_{m}^{b}+1,Q_{l}^{a},Q_{l}^{b}\big)\quad&& \mbox{if } x<\min\left\{-\frac{\epsilon_{m}}{2k}-1-\bar x,\frac{\epsilon_{l}-2k-\epsilon_{m}}{2k(1-p)}-\frac{p}{2}-\bar x\right\}, \notag\\
&LB\!&&:&&V\big(Q_{m}^{a},Q_{m}^{b},Q_{l}^{a},Q_{l}^{b}+1\big)\quad&&\mbox{if } \frac{\epsilon_{l}-2k-\epsilon_{m}}{2k(1-p)}-\frac{p}{2}-\bar x<x<-\frac{\epsilon_{l}}{2kp}+\frac{1-p}{2}-\bar x,\notag \\ 
&DP\!&&:&&V\big(Q_{m}^{a},Q_{m}^{b},Q_{l}^{a},Q_{l}^{b}\big)\quad&& \mbox{if }\max\left\{-\frac{\epsilon_{m}}{2k}-1-\bar x,-\frac{\epsilon_{l}}{2kp}+\frac{1-p}{2}-\bar x\right\} \notag\\
& && && && \,\,\,\,\,\, \leq x\leq\min\left\{\frac{\epsilon_{m}}{2k}+1-\bar x,\frac{\epsilon_{l}}{2kp}-\frac{1-p}{2}-\bar x\right\},  \notag\\ 
&LS\!&&:&&V\big(Q_{m}^{a},Q_{m}^{b},Q_{l}^{a}+1,Q_{l}^{b}\big)\quad&&\mbox{if }\frac{\epsilon_{l}}{2kp}-\frac{1-p}{2}-\bar x<x<\frac{-\epsilon_{l}+2k+\epsilon_{m}}{2k(1-p)}+\frac{p}{2}-\bar x,\notag \\ 
&MS\!&&:&&V\big(Q_{m}^{a}+1,Q_{m}^{b},Q_{l}^{a},Q_{l}^{b}\big)\quad&&\mbox{if } x>\max\left\{\frac{\epsilon_{m}}{2k}+1-\bar x,\frac{-\epsilon_{l}+2k+\epsilon_{m}}{2k(1-p)}+\frac{p}{2}-\bar x\right\},
\end{alignedat}\right.
$$
\\
where $
\bar x=\big(Q_{l}^{b}-Q_{l}^{a}\big)p+Q_{m}^{b}-Q_{m}^{a}+c_{2}(T-n\Delta).
$
Limit buy orders and limit sell orders are never the optimal choice if $\epsilon_{l}<p(\epsilon_{m}+3k+pk)$. Take for example limit buy orders. At each time $t_{i}$, they are the optimal choice if and only if the inventory $x$ satisfies
\begin{equation}
\frac{\epsilon_{l}-2k-\epsilon_{m}}{2k(1-p)}-\frac{p}{2}-\bar x<x<-\frac{\epsilon_{l}}{2kp}+\frac{1-p}{2}-\bar x.
\label{eq:invcond}
\end{equation}
For such $x$ to exist, one must have 
$$
\frac{\epsilon_{l}-2k-\epsilon_{m}}{2k(1-p)}<\frac{-\epsilon_{l}+kp}{2kp}\Leftrightarrow0<\epsilon_{l}<p(\epsilon_{m}+3k+pk).
$$
The same condition also applies for the existence of limit sell orders. Since the above condition neither depends on  time, nor on the strategy followed thus far, we may regard it as an intertemporal and strategy-independent condition. Following an analogous procedure, we find that the above conditions ensure the existence of limit sell orders, too.
\noindent In Figure \ref{eq:vf1}, we plot the value function obtained by  the recursive relation given above. In  Figure \ref{eq:vf23}, we show that the condition $\epsilon_{l}<p(\epsilon_{m}+3k+pk)$ ensures that limit orders are the optimal choice when the inventory satisfies Equation (\ref{eq:invcond}). In particular, this means that limit orders are optimal if the penalty for trading in the lit pool by means of limit orders does not exceed the quantity $p(\epsilon_{m}+3k+pk)$ as in Figure \ref{eq:vf23} (a). On the contrary, in Figure \ref{eq:vf23} (b) we note that an optimal combination of market orders and dark pool activity outperforms limit orders and thus the latter are never the optimal choice.

\noindent We now remove assumptions (a), (b),  (f), (g), and (h) and replace them with
\begin{enumerate}[(a')]
\item the mid-price process follows a geometric Brownian motion,
\item the dark-pool orders $a$ and $b$  are of random size with supports $A$ and $B$ respectively,
\end{enumerate}
\begin{enumerate}[(f')]
\item the limit price at which the dark pool posts in the lit pool can be optimally chosen, i.e. $\kappa\neq0$,
\end{enumerate}
\begin{enumerate}[(g')]
\item the dark pool can observe and control its cash and inventory levels on a continuous-time basis,
\end{enumerate}
\begin{enumerate}[(h')]
\item $g(u,x)=-\phi g(x)=-\phi x^{2}=$ and $U(x,y,s)=y+x(s-k|x|/x)$. That is, the dark pool is subject to a quadratic running penalty for holding a non-zero inventory and liquidates all its inventory at the final date $T$. This manifests the dark pool's risk-adversity since a large inventory reduces the value function of the pool.
\end{enumerate}
 We assume that $\PR[z=0]=\ell^{\kappa}(z_{0})$ and  $\PR[z=1]=\ell^{\kappa}(z_{1})=1-\ell^{\kappa}(z_{0})$. This reflects the fact that the filling-probability of a limit order depends on how far from the mid-price such an order is posted. The  associated QVI is
\begin{equation*}
  \begin{alignedat}{4}
  &{\hspace{-2cm}}  \min\Bigg\{ \hspace{-0.4cm} \quad \phi g(x)-\partial_{t} V&&{\hspace{-11.8cm}}-\frac{1}{2}\sigma^{2}s^{2}\partial_{ss}V  \\
    & &&{\hspace{-11.8cm}} -\lambda^{a}\int_{A}\!\!\Big[V(t,x-a,y+a(s+\delta^{a}),s)-V(t,x,y,s)\Big]\ell_{a}(a)\rd a \\
     &\hspace{-1.8cm}  \quad  && {\hspace{-11.8cm}} -\lambda^{b}\int_{B}\!\!\Big[V(t,x+b,y-b(s-\delta^{b}),s)-V(t,x,y,s)\Big]\ell_{b}(b)\rd b;\\
     &\hspace{-1.8cm}  \quad V(t,x,y,s)-\sup_{\xi=\pm1}\left[V\left(t,x+\xi,y-\xi\left(s+\xi k\right),s\right)-\epsilon_{m}\right]; & \quad   \\
     & \hspace{-1.8cm} \quad V(t,x,y,s)-\sup_{\eta=\pm1,\kappa\in[0,\bar\kappa]}\sum_{i=0}^{1}\left[V\left(t,x+\eta z_{i},y-z_{i}\left(s-\eta\left( k+\kappa\right)\right),s\right)-\epsilon_{l}\right]\ell^{\kappa}(z_{i})\Bigg\}=0.& \quad 
  \end{alignedat}
\end{equation*}
\newline
Given the form of the terminal condition, we consider the ansatz $V(t,x,y,s)=y+xs+h(t,x)$
with terminal condition $h(T,x)=-k|x|-\epsilon_{m}$. We refer to Cartea \& Jaimungal (2012) and Cartea et al. (2014) for more details about this ansatz. We then get:\\
\newline
\begin{equation*}
\begin{alignedat}{4}
\min\Bigg\{&\phi g(x)-\partial_{t} h(t,x)&&\hspace{-8.4cm}-\lambda^{a}\int_{A}\Big[a\delta^{a}+h(t,x-a)-h(t,x)\Big]\ell_{a}(a)\rd a\\
& &&\hspace{-8.4cm}-\lambda^{b}\int_{B}\Big[b\delta^{b}+h(t,x+b)-h(t,x)\Big]\ell_{b}(b)\rd b;\\
&h(t,x)-\sup_{\xi=\pm1}\left[-|\xi| k-\epsilon_{m}+h(t,x+\xi)\right];\\
&h(t,x)-\sup_{\eta=\pm1,\kappa\in[0,\bar\kappa]}\sum_{i=0}^{1}\left[|\eta|\left( k+\kappa\right)z_{i}-\epsilon_{l}+h(t,x+\eta z_{i})\right]\ell^{\kappa}(z_{i})\Bigg\}=0.
\end{alignedat}
\end{equation*}

\begin{figure}[H]
\begin{center}
\includegraphics[width=350pt,height=180pt]{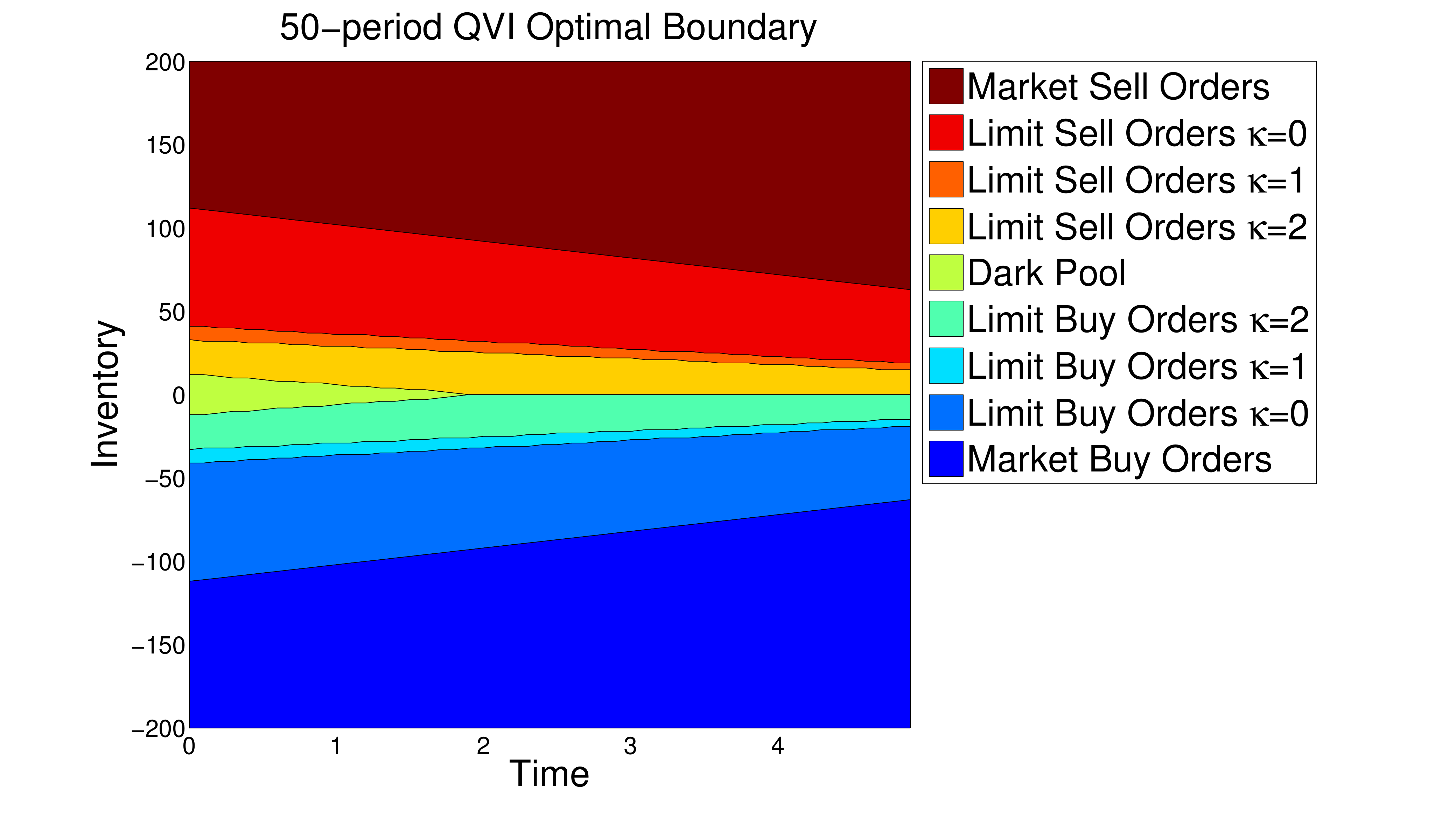}
\caption{\footnotesize{Optimal boundary for the doube-obstacle problem
 with $\kappa=0,1,2$.}}
\label{eq:g1}
\end{center}
\end{figure}
\begin{figure}[H]
\begin{center}
\includegraphics[width=350pt,height=160pt]{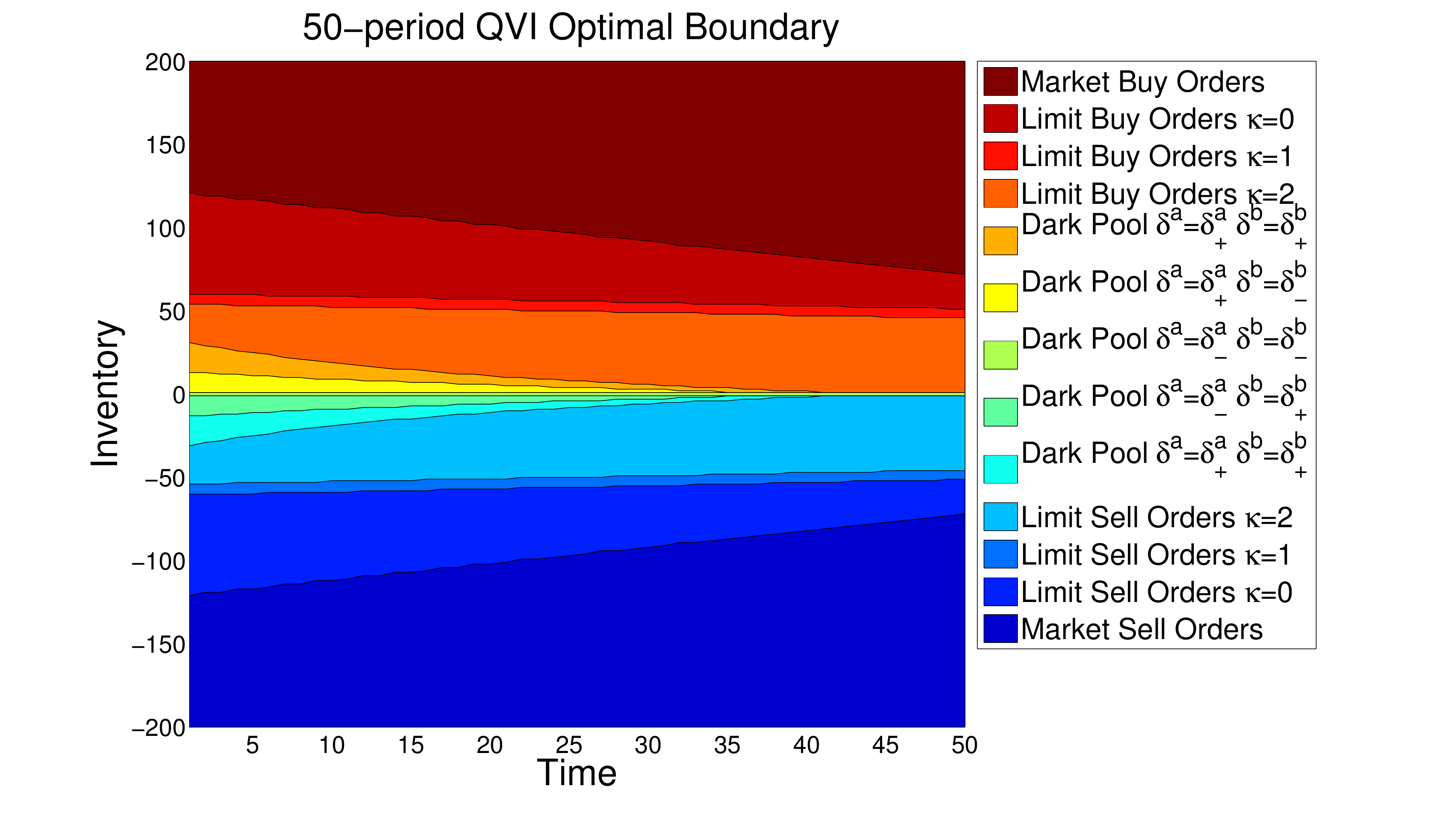}
\caption{\footnotesize{Optimal boundary for the doube-obstacle problem
 with $\kappa=0,1,2$ and $\delta^{a}_{-}=\delta^{b}_{-}=0.2$ and $\delta^{a}_{+}=\delta^{b}_{+}=0.4$.
}}
\label{eq:g2}
\end{center}
\end{figure}

\vspace{-0.5cm}
 \noindent As one would expect, as the inventory increases, the dark pool will first resort to limit orders and ultimately to market orders. We  note that when posting a limit order, the smaller the inventory, the farther from the mid-price the market maker posts. A pre-specified non-zero inventory would produce a shift in the optimal boundaries by an equal amount. Moreover, the symmetry in Figure \ref{eq:g1} is due to the particular choice made for the frequency of orders submitted by the clients to the dark pool. That is, by choosing $\lambda^{a}\neq\lambda^{b}$, we would loose such a symmetry and see a shift upwards (resp. downwards) of the optimal boundaries when $\lambda^{a}<\lambda^{b}$ (resp. ($\lambda^{a}>\lambda^{b}$). To solve the QVI, finite difference methods were applied. We now replace assumption (d) with
\begin{enumerate}[(d')]
\item the dark pool optimally chooses the commissions $\delta^{a}\in\big\{\delta^{a}_{-},\delta^{a}_{+}\big\}$ and $\delta^{b}\in\big\{\delta^{b}_{-},\delta^{b}_{+}\big\}$.
\end{enumerate}
The  QVI now reads:
\begin{equation}
\begin{alignedat}{4}
\min\Bigg\{&\phi g(x)-\partial_{t} h(t,x)&&\hspace{-8.5cm}-\sup_{\delta^{a}=\delta^{a}_{\pm}}\lambda^{a}(\delta^{a})\int_{A}\!\!\Big[a\delta^{a}+h(t,x-a)-h(t,x)\Big]\ell_{a}(a)\rd a\\
& &&\hspace{-8.5cm}-\sup_{\delta^{b}=\delta^{b}_{\pm}}\lambda^{b}(\delta^{b})\int_{B}\Big[b\delta^{b}+h(t,x+b)-h(t,x)\Big]\ell_{b}(b)\rd b;\\
&h(t,x)-\sup_{\xi=\pm1}\left[-|\xi| k-\epsilon_{m}+h(t,x+\xi)\right];\\
&h(t,x)-\sup_{\eta=\pm1,\kappa\in[0,\bar\kappa]}\sum_{i=0}^{1}\left[|\eta|\left( k+\kappa\right)z_{i}-\epsilon_{l}+h(t,x+\eta z_{i})\right]\ell^{\kappa}(z_{i})\Bigg\}=0.
\label{eq:sim2}
\end{alignedat}
\end{equation}
In Figure \ref{eq:g2} we plot the numerical solution of $(\ref{eq:sim2})$.
\vspace{-0.5cm}
\noindent In Figure \ref{eq:g2} we see that when the inventory is relatively small, it is optimal to set $\delta^{a}=\delta^{a}_{-}$ and $\delta^{b}=\delta^{b}_{-}$ so to encourage the arrival of dark pool orders. Then, for example,  when  the inventory increases, the incentive is lowered to $\delta^{a}=\delta^{a}_{+}$ and $\delta^{b}=\delta^{b}_{-}$.  If the inventory  increases further, $\delta^{a}=\delta^{a}_{+}$ and $\delta^{b}=\delta^{b}_{+}$ is preferred since the dark pool must increase the commissions to avoid reaching the threshold at which it will need to place orders in the lit pool. The critical inventory level at which the dark pool begins placing orders in the lit pool falls as the terminal liquidation date is approached. Finally, we observe that the situation is symmetric when the inventory is negative, i.e. the commissions increase as the inventory decreases and the lit pool thresholds shrink near the terminal date. In the last simulation, we remove assumption (i) and replace it with
\begin{enumerate}[(i')]
\item a stochastic bid-ask spread is introduced.
\end{enumerate}
For the ansatz $V(t,x,y,s;k)=y+xs+h_{k}(t,x)$, the system of  QVIs can be reduced to
\begin{equation}
\begin{alignedat}{4}
&\min\Bigg\{\phi g(x)-\partial_{t} h_{k}(t,x)\ -\sup_{\delta^{a}=\delta^{a}_{\pm}}\lambda^{a}(\delta^{a})\int_{A}\!\!\Big[a\delta^{a}+h_{k}(t,x-a)-h_{k}(t,x)\Big]\ell_{a}(a)\rd a\\
&-\sup_{\delta^{b}=\delta^{b}_{\pm}}\lambda^{b}(\delta^{b})\int_{B}\Big[b\delta^{b}+h_{k}(t,x+b)-h_{k}(t,x)\Big]\ell_{b}(b)\rd b-\sum_{j\neq k}r_{kj}\Big[h_{j}(t,x)-h_{k}(t,x)\Big];\\
&h_{k}(t,x)-\sup_{\xi}\left[-|\xi| k-\epsilon_{m}+h_{k}(t,x+\xi)\right];\\
&h_{k}(t,x)-\sup_{\eta,\kappa\in[0,\bar\kappa]}\sum_{i=0}^{1}\left[|\eta|\left( k+\kappa\right)z_{i}-\epsilon_{l}+h_{k}(t,x+\eta z_{i})\right]\ell^{\kappa}(z_{i})\Bigg\}=0.
\label{eq:sim4}
\end{alignedat}
\end{equation}
\begin{figure}[H]
\begin{center}
\hspace{-0.5cm}\subcaptionbox{\footnotesize{1st regime with $k=0.8$}\label{fig:1a}}{\includegraphics[width=200pt,height=150pt]{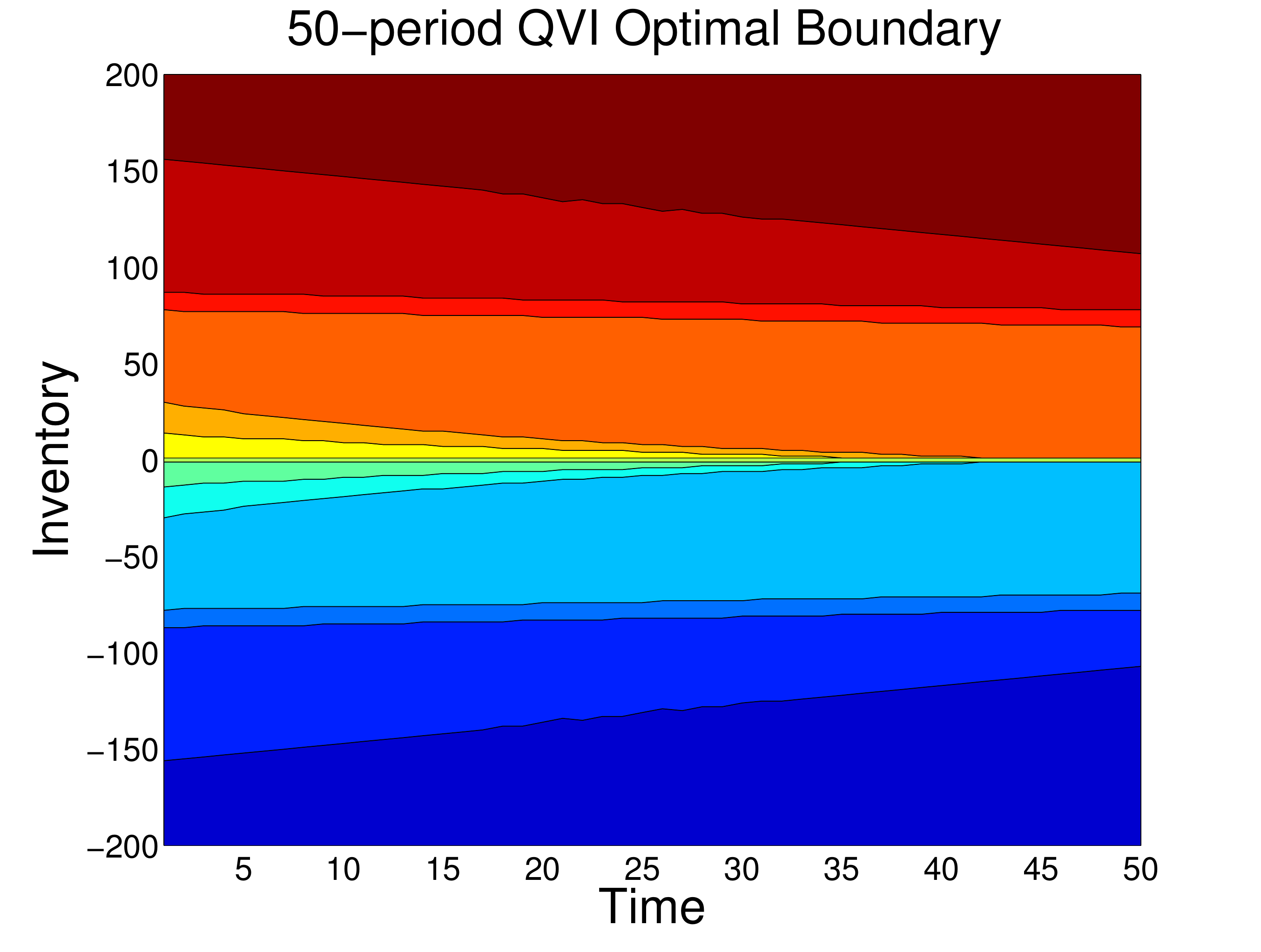}}
\hspace{-0.5cm}\subcaptionbox{\footnotesize{2nd regime with $k=1.3$}\label{fig:1b}}{\includegraphics[width=200pt,height=150pt]{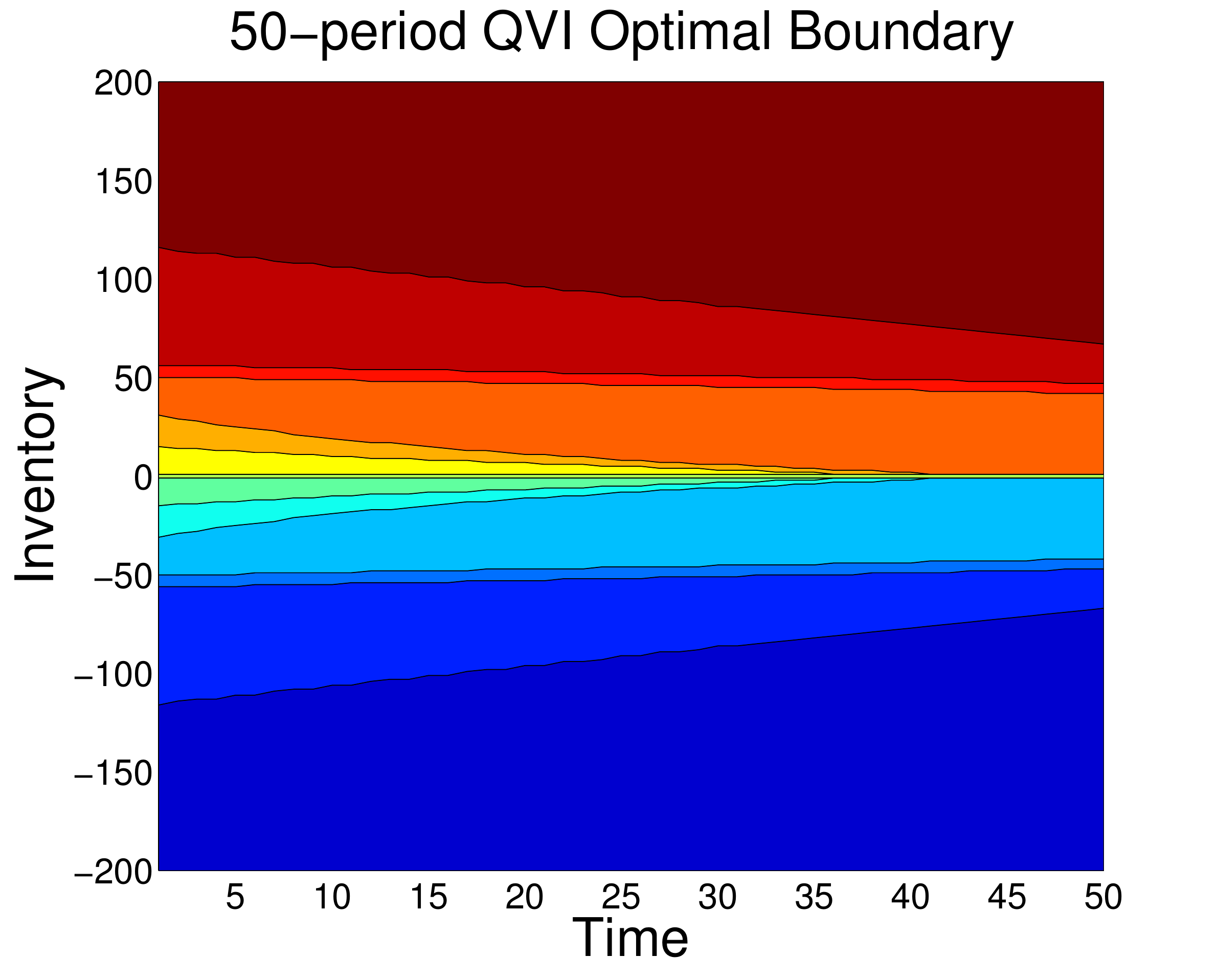}}
\subcaptionbox{\footnotesize{3rd regime with $k=1.8$}\label{fig:1c}}{\includegraphics[width=330pt,height=150pt]{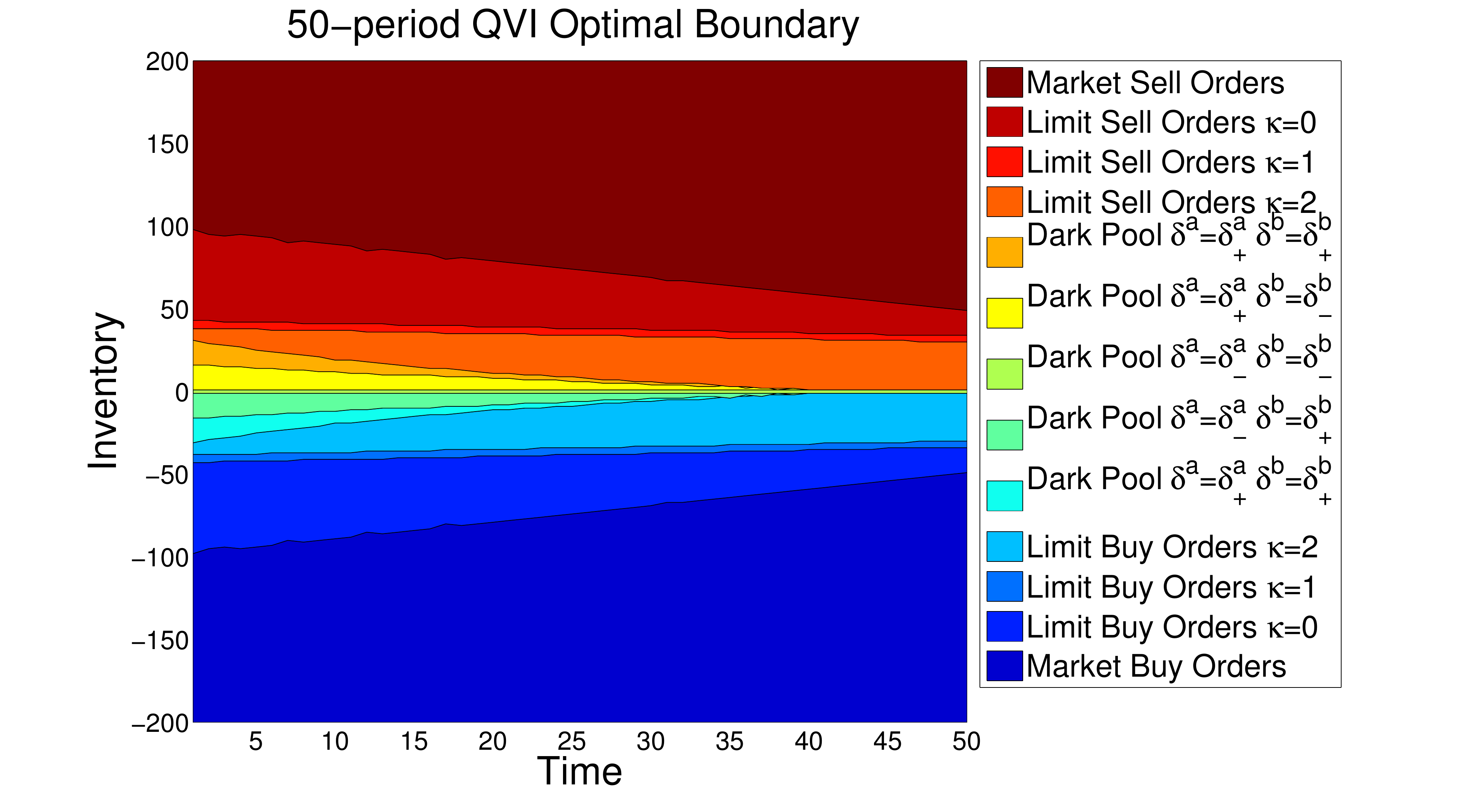}}
\caption{\footnotesize{Optimal boundary for the doube-obstacle problem
 with $\kappa=0,1,2$ and $\delta^{a}_{-}=\delta^{b}_{-}=0.2$, $\delta^{a}_{+}=\delta^{b}_{+}=0.4$ and $k=0.8,1.3,1.8$.}}
\label{eq:g33}
 \end{center}
\end{figure}
\noindent In Figure \ref{eq:g33} we note that as the spread increases, the lit pool thresholds decrease. In fact, the market maker tends to resort  to the lit pool earlier in order to avoid holding a large inventory at time $T$, which will indeed be very expensive to dispose of.
\section{Concluding remarks}
In the present work we study an optimal market-making problem faced by a dark pool. The dark pool earns the commission fee from executing the trade orders within the pool placed by its clients. The market agents, who trade through the dark pools, benefit from anonymity and advantageous prices. Throughout the activity the pool faces an inventory risk, which can be reduced (i) by controlling the width of the dark pool spread, and (ii) by resorting to the lit pool via both, market and limit orders. The dark pool activity is preferred since it protects from   information leakage. Such a feature is modelled via a fixed penalty incurred by the dark pool whenever it submits an order to the lit pool. As confirmed by the numerical results, the dark pool will refrain from placing orders in a lit pool as long as the size of the inventory  is relatively small. Whenever a certain level is exceeded, the dark pool resorts to the lit pool by means of limit orders. A limit order is more remunerative though its execution is uncertain. The dark pool can choose the limit price; we find that the more the inventory grows, the closer to the mid-price the dark pool will post. This is reasonable since  the filling-intensity of limit orders depends on how far from the mid-price they are posted. If the inventory becomes critically large, market orders will be preferred instead, which are costly but benefit of sure execution. When the end of the market-making activity approaches (which, e.g., might be thought of as the end of the trading day) the market-orders region in the lit pool  widens while the dark-pool and limit-orders regions in the lit pool diminish. In fact, the market maker will incur in a higher penalty by liquidating a large inventory at the terminal date. These conclusions are obtained by formulating a double obstacle standard stochastic and impulse control problem, and  by showing that the associated value function is the unique  viscosity solution of the corresponding dynamic programming equation. Finally, we provide four numerical examples with increasing complexity. The  numerical analysis is similar to the one carried out by Guilbaud \& Pham (2013) and our results concerning the lit pool are in line with the ones obtained by them. 

\section{Appendix}
Here state all the standing assumptions which are introduced in the modelling setup presented in this paper.
\begin{enumerate}[(i)]
\item  the map $\mu:[0,T]\times \RR_{+} \rightarrow \RR$ and $\sigma:[0,T]\times \RR_+ \rightarrow \RR_+$ satisfy Lipschitz continuity 
\begin{equation*}
\begin{split}
\big|\mu\left(t_1,s_1\right)-\mu\left(t_2,s_2\right)\big|^{2}+\left|\sigma\left(t_1,s_1\right)-\sigma\left(t_2,s_2\right)\right|^{2}&\leq C\left(|t_1-t_2|^{2}+|s_1-s_2|^{2}\right)\\
\end{split}
\end{equation*}
for all $t_1$, $t_2\in[0,T]$ and $s_1,s_2\in\RR_{+}$;
\item $f:[0,T]\times\big[\underline{X}\,,\overline{X}\big]\times\RR_+\cup\{0\} \rightarrow \RR_+\cup\{0\}$  satisfies Lipschitz continuity and the linear growth conditions
\begin{equation*}
\begin{split}
\left|f\left(t_1,x_1,v\right)-f\left(t_2,x_2,v\right)\right|^{2}&\leq C\left(|t_1-t_2|^{2}+|x_1-x_2|^{2}\right)\\
\left|f\left(t_1,x_1,v\right)\right|^{2}&\leq C\left(1+|x_1|^{2}\right)\end{split}
\end{equation*}
for all $t_1$, $t_2\in[0,T]$, $x_1$, $x_2\in\big[\underline{X}\,,\overline{X}\big]$, and $v\in\RR$;
\item $f_1:[0,T]\times \mathcal S\times[0,k]\times\RR_+\cup\{0\} \rightarrow  \RR_+\cup\{0\}$  satisfies Lipschitz continuity and the linear growth conditions
\begin{equation*}
\begin{split}
\left|f_1\left(t_1,{\boldsymbol x}_1,d,v\right)-f_1\left(t_2,{\boldsymbol  x}_2,d,v\right)\right|^{2}&\leq C\left(|t_1-t_2|^{2}+\|{\boldsymbol  x}_1-{\boldsymbol x}_2\|_2^{2}\right)\\
\left|f_1\left(t_1,{\boldsymbol  x}_1,d,v\right)\right|^{2}&\leq C\left(1+\|{\boldsymbol  x}_1\|_2^{2}\right)\end{split}
\end{equation*}
for all $t_1$, $t_2\in[0,T]$, $d\in[0,k]$ and ${\boldsymbol  x}_1$, ${\boldsymbol  x}_2 \in\mathcal S$;
\item $\Gamma:\mathcal N\times\big[\underline{X}\,,\overline{X}\big]\times[0,1]\rightarrow\RR$ and $\chi:\mathcal N\times\big[\underline{X}\,,\overline{X}\big]\times[0,1]\times \RR_{+}\times[0,\bar \kappa]\times\mathbb K\rightarrow\RR$ are Lipschitz continuous functions satisfying, for $M>0$, the following properties:
\begin{enumerate}[(a)]
\item $0\leq\Gamma(\eta,x,z)\leq M$ and $\chi(\eta,x,z,s,k,\kappa)\leq 0$ if $x\leq0$, for all $\eta\in\mathcal N$, $z\in[0,1]$,  $(s,\kappa)\in\RR_{+}\times[0,\bar\kappa]$ and $k\in\mathbb K$,
\item $-M\leq\Gamma(\eta,x,z)\leq 0$ and $\chi(\eta,x,z,s,k,\kappa)\geq 0$ if $x\geq0$, for all $\eta\in\mathcal N$, $z\in[0,1]$,  $(s,\kappa)\in\RR_{+}\times[0,\bar\kappa]$ and $k\in\mathbb K$;
\item $|x|^{2}-\E\big[|x+\Gamma(\eta,x,z)|^{2}\big]>1$, for all $\eta\in\mathcal N$, $z\in[0,1]$ and $x\in\big[\underline{X}\,,\overline{X}\big]$;
\end{enumerate}
\item $\Lambda:\mathcal X\times\big[\underline{X}\,,\overline{X}\big]\rightarrow\RR$ and $c:\mathcal X\times\big[\underline{X}\,,\overline{X}\big]\times \RR_{+}\times\mathbb K\rightarrow\RR$ are Lipschitz continuous  functions satisfying, for $M>0$, the following properties:
\begin{enumerate}[(a)]
\item $0\leq\Lambda(\xi,x)\leq M$ and $ c(\xi,x,s,k)\leq 0$ if $x\leq0$, for all $\xi\in\mathcal X$, $s\in\RR_{+}$ and $k\in\mathbb K$,
\item $-M\leq\Lambda(\xi,x)\leq 0$ and $c(\xi,x,s,k)\geq 0$ if $x\geq0$, for all $\xi\in\mathcal X$, $s\in\RR_{+}$ and $k\in\mathbb K$;
\item $|x|^{2}-|x+\Lambda(\xi,x)|^{2}>1$, for all $\xi\in\mathcal X$ and $x\in\big[\underline{X}\,,\overline{X}\big]$;
\end{enumerate}
\item $g:[0,T]\times\big[\underline{X}\,,\overline{X}\big]\rightarrow\RR$ satisfies Lipschitz continuity 
\begin{equation*}
\begin{split}
\left|g\left(t_1,x_1\right)-g\left(t_2,x_2\right)\right|^{2}&\leq C\left(|t_1-t_2|^{2}+|x_1-x_2|^{2}\right)\\
\end{split}
\end{equation*}
for all $t_1$, $t_2\in[0,T]$ and  $x_1$, $x_2\in\big[\underline{X}\,,\overline{X}\big]$;
\item $U:\mathcal S\times\mathbb K\rightarrow\RR$ satisfies Lipschitz continuity and the linear growth conditions
\begin{equation*}
\begin{split}
\left|U\left({\boldsymbol x}_1,k\right)-U\left({\boldsymbol x}_2,k\right)\right|^{2}&\leq C\left(\|{\boldsymbol x}_1-{\boldsymbol x}_2\|_2^{2}\right)\\
\left|U\left({\boldsymbol x}_1,k\right)\right|^{2}&\leq C\left(1+\|{\boldsymbol x}_1\|_2^{2}\right),
\end{split}
\end{equation*}
for all ${\boldsymbol x}_1$, ${\boldsymbol x}_2\in\mathcal S$ and $k\in\mathbb K$.
\end{enumerate}

\subsection{Viscosity Solution \ref{eq:visc}}
\begin{prop}(Existence)
{\itshape The system of functions $V(t,{\bf x};k)$ is a  viscosity solution of the QVI (\ref{eq:qvi}).
\label{eq:visc}}
\end{prop}
\noindent{\itshape Proof.} We use definition \ref{eq:def1} and we show that the system of functions $V(t,{\bf x};k)$ is a viscosity solution by proving that it is both a supersolution and a subsolution. First we note that we have $V(T,{\boldsymbol 
x};k)=U({\boldsymbol x};k)$ on $\{T\}\times \mathcal S\times\mathbb K$, thus we need to prove the viscosity property only on $[t,T)\times \mathcal S\times\mathbb K$.
Results in, e.g., Ly Vath et al. (2007) ensure that $\mathcal M V_{*}\leq (\mathcal MV)_{*}$ and $\mathcal L V_{*}\leq (\mathcal LV)_{*}$. By definition of the value function, we have $V\geq\mathcal MV$ and $V\geq \mathcal L V$ for all $(u,{\bf x})\in[t,T)\times\mathcal S$. It follows that  $V_{*}\geq(\mathcal MV)_{*}\geq\mathcal M V_{*}$ and    $V_{*}\geq (\mathcal LV)_{*}\geq\mathcal LV_{*}$. That is, it suffices to show that 
$$
-g(\bar t,\bar x)\!- \mathcal A\left(\bar t,\bar{\boldsymbol x},\bar k,\partial_{t}\phi,\partial_{s}\phi,\partial_{ss}\phi,\phi\right) \geq0.
$$
Let $(V_{*}-\phi)(\bar t,\bar{\boldsymbol x};\bar k)=0$, where $(\bar t,\bar{\boldsymbol x},\bar k)=\arg\min (V_{*}-\phi)( t, {\boldsymbol x};k)$. By definition of $V_{*}$, there exists a sequence $(t_{m},{\boldsymbol x}_{m})\rightarrow (\bar t,\bar{\boldsymbol x})$ such that $V_{*}(t_{m},{\boldsymbol x}_{m};\bar k)\rightarrow V_{*}(\bar t, \bar{\boldsymbol x};\bar k)$ as $m\rightarrow\infty$. We define the stopping time
\begin{equation}
\theta_{m}=\inf\left\{u> t_{m}\,|\,{\boldsymbol X}_{t_{m}\,{\boldsymbol x}_{m}}(u)\notin B_{\eta}( t_{m},{\boldsymbol x}_{m})\right\},
\label{eq:st}
\end{equation}
 where $B_{\eta}(t_{m},{\boldsymbol x}_{m})$ is the open ball of radius $\eta$ centred in $( t_{m}, {\boldsymbol x}_{m})$. We choose a strictly positive sequence $h_{m}\rightarrow0$ and let the stopping time $\theta_{m}^{*}:=\theta_{m}\wedge( t_{m}+h_{m})\wedge\theta^{*}\wedge \tau_m$, where $\theta^{*}$ is the first time the regime switches from its initial value $\bar k$ and where $\tau_m$ is the first time an impulse takes place. By the first part of the DPP and the definition of the function $\phi$, we have for any admissible control strategy
$$
V(t_{m},{\boldsymbol x}_{m};\bar k)\geq\E\Biggl[\int_{ t_{m}}^{\theta_{m}^{*}}g\big(u,X_{ t_{m}, \, x_m}(u)\big)\rd u + \phi\big(\theta^{*}_{m},{\boldsymbol X}_{t_{m},\,{\boldsymbol x}_{m}}\big(\theta^*_{m}\big);k\big(\theta^*_{m}\big)\big)\Biggr].
$$
An application of It$\oo$'s formula to $\phi$ between $t_{m}$ and $\theta_{m}^{*}$ yields
\begin{equation*}
\begin{split}
(V_{*}-\phi)\big(t_{m},{\boldsymbol x}_{m};\bar k\big)\geq\E\Biggl[\int_{ t_{m}}^{\theta_{m}^{*}}\Big[g\big(u,&\ X_{ t_{m}, \, x_m}(u)\big)\\
&+ \bar{\mathcal A}\big(u ,{\boldsymbol X}_{t_{m},\,{\boldsymbol x}_{m}}(u),\bar k,\partial_{t}\phi,\partial_{s}\phi,\partial_{ss}\phi,V_{*},D\big)\Big]  \rd u\Biggr].
\end{split}
\end{equation*}
We can divide by $-h_{m}$,  then let $m\rightarrow\infty$  and apply the  the mean value theorem. Finally, the result follows form the arbitrariness of the control variable. 
First note that if $V^{*}\leq\mathcal M V^{*}$ or $V^{*}\leq \mathcal LV^{*}$, the subsolution property is immediately satisfied. We assume therefore that  $V^{*}>\mathcal M V^{*}$ and $V^{*}> \mathcal LV^{*}$;  we then need to show that 
\begin{equation}
-g(\bar t,\bar x)\!- \mathcal A\left(\bar t,\bar{\boldsymbol x},\bar k,\partial_{t}\phi,\partial_{s}\phi,\partial_{ss}\phi,\phi\right) \leq0.
\label{eq:gaga}
\end{equation}
By continuity of the mapping in (\ref{eq:gaga}), we assume on the contrary that there exist a $\delta>0$ and an $\epsilon>0$ such that $-g(\bar t,\bar x)\!- \mathcal A\left(\bar t,\bar{\boldsymbol x},\bar k,\partial_{t}\phi,\partial_{s}\phi,\partial_{ss}\phi,\phi\right)\geq \delta$, for all ${\boldsymbol X}_{\bar t,\,\bar{\boldsymbol  x}}(u)\in B_{\epsilon}(\bar t,\bar{\boldsymbol x})$. We take the sequences $h_{m}\rightarrow0$ and  $(t_{m},{\boldsymbol x}_{m})\rightarrow (\bar t,\bar{\boldsymbol x})$ valued in $B_{\epsilon}(\bar t,\bar{\boldsymbol x})$ and we define the stopping times $\theta_{m}$ by  (\ref{eq:st}) with $\eta<\epsilon$ and $\theta_{m}^{*}:=\theta_{m}\wedge( t_{m}+h_{m})\wedge\theta^{*}\wedge\tau_m$. By  It\^{o}'s formula and  the second part of the DPP\,, there exists an admissible control strategy $D^{*}$ for which
\begin{equation*}
\begin{split}
\gamma_{m}-\frac{\delta h_{m}}{2}&\leq\E\Biggl[\int_{ t_{m}}^{\theta_{m}^{*}}\left[g\big(u,X_{ t_{m}, \, x_m}(u)\big)+ \bar{\mathcal A}\big(u ,{\boldsymbol X}_{t_{m},\,{\boldsymbol x}_{m}}(u),\bar k,\partial_{t}\phi,\partial_{s}\phi,\partial_{ss}\phi,\phi,D^{*}\big)\right]  \rd u\Biggr],
\end{split}
\end{equation*}
where $\gamma_{m}=(V^{*}-\phi)\big(t_{m},{\boldsymbol x}_{m};\bar k\big)$. Dividing by $-h_{m}$, we find that
$$
0\geq\frac{\gamma_{m}}{h_{m}}-\frac{\delta}{2}+\frac{\delta}{h_{m}}\E[\theta^{*}_{m}-t_{m}].
$$
Since $\E[\theta^{*}_{m}-t_{m}]\delta/h_{m}\rightarrow1$ as $m\rightarrow\infty$, we get $\delta/2\leq0$, which contradicts  $\delta>0$.\hfill$\square$
\subsection{Strong Comparison Results \ref{eq:unique}}
\begin{prop}(Strong Comparison Principle)
{\itshape Let $ V$ and $U$ be a supersolution and a subsolution respectively  of the QVI (\ref{eq:qvi}). If  $U(T,\cdot)\leq V(T,\cdot)$, then $U\leq V$ on $[0,T]\times \mathcal S\times \mathbb K$.}
\label{eq:unique}
\end{prop}
\vspace{-0.5cm}
\noindent{\itshape Proof.} Let $v=v_{*}$ and $u=u^{*}$ be a supersolution and a subsolution respectively. We first prove that there exists a $\zeta$-strict supersolution, where $0<\zeta<\epsilon_l$. We refer to, e.g., Seydel (2009) for technical details. We consider the function $v^{\zeta}(t,{\boldsymbol x};k)=v(t,{\boldsymbol x};k)+\zeta\e^{\beta(T-t)}(1+|x|^{2p})$, where $\beta>0$ and $p>1$ are to be determined later. Then we have:
\begin{equation*}
\begin{split}
&v^{\zeta}(t,{\boldsymbol x};k)-\mathcal Mv^{\zeta}(t,{\boldsymbol  x};k)\\
&\geq v(t,{\boldsymbol x};k)+\zeta\e^{\beta(T-t)}(1+|x|^{2p})-\mathcal Mv(t,{\boldsymbol x};k)\\
&\hspace{0.5cm}-\sup_{\xi\in\mathcal X}\left[\zeta\e^{\beta(T-t)}\left(1+|x+\Lambda(\xi,x)|^{2p}\right)-\epsilon_m\right]\\
&\geq\zeta\e^{\beta(T-t)}\left[|x|^{2p}-  \sup_{\xi\in\mathcal X}\left(|x+\Lambda(\xi,x)|^{2p}\right)\right]+\epsilon_m,  \\
\end{split}
\end{equation*}
where the last inequality follows form the supersolution property of the function $v$. Furthermore, by assumption (v) for the function $\Lambda$, we have 
$$
\zeta\e^{\beta(T-t)}\left[|x|^{2p}-  \sup_{\xi\in\mathcal X}\left(|x+\Lambda(\xi,x)|^{2p}\right)\right]+\epsilon_m>\zeta,
$$
since $|a|>|b|\Rightarrow |a|^p>|b|^p\forall \,p>1$. Analogously, we have:
\begin{equation*}
\begin{split}
&v^{\zeta}(t,{\boldsymbol x};k)-\mathcal Lv^{\zeta}(t,{\boldsymbol  x};k)\\
&\geq \zeta\e^{\beta(T-t)}\left[|x|^{2p}- \sup_{\eta\in\mathcal N,\,\kappa\in[0,\bar\kappa]}\int_0^1\left(|x+\Gamma(\eta,x,z)|^{2p}\right)\ell^{(\kappa)}_{z}(z)\rd z\right]>\zeta.
\end{split}
\end{equation*}
 Finally we take into consideration  the PIDE part. We let $\phi^{\zeta}$ be the test function for $v^{\zeta}$. Then $\phi:=\phi^{\zeta}-\zeta\e^{\beta(T-t)}(1+|x|^{2p})$ is the test function for $v$. We therefore have:\\
\begin{equation*}
\begin{split}
& -g(t,x)-\mathcal A\left( t,{\boldsymbol  x},k,\partial_{t}\phi^{\zeta},\partial_{s}\phi^{\zeta},\partial_{ss}\phi^{\zeta},v^{\zeta}\right)\\
&\geq -g(t,x)-\mathcal A\left( t,{\boldsymbol  x},k,\partial_{t}\phi,\partial_{s}\phi,\partial_{ss}\phi,v\right)+\beta\zeta\e^{\beta(T-t)}(1+|x|^{2p})\\
&\hspace{0.5cm}-\sup_{\delta^{a}\in[0,k]}\lambda^a_{\delta}\int_0^\infty\zeta\e^{\beta(T-t)}\left(|x+f(t,x,a)|^{2p}-|x|^{2p}\right)\ell_{a}(a)\rd a\\
&\hspace{0.5cm}-\sup_{\delta^{b}\in[0,k]}\lambda^b_{\delta}\int_0^\infty\zeta\e^{\beta(T-t)}\left(|x-f(t,x,b)|^{2p}-|x|^{2p}\right)\ell_{b}(b)\rd b,\\
\end{split}
\end{equation*}
which implies that
\begin{equation*}
\begin{split}
& -g(t,x)-\mathcal A\left( t,{\boldsymbol  x},k,\partial_{t}\phi^{\zeta},\partial_{s}\phi^{\zeta},\partial_{ss}\phi^{\zeta},v^{\zeta}\right)\\
&\geq \beta\zeta\e^{\beta(T-t)}(1+|x|^{2p})-\sup_{\delta^{a}\in[0,k]}\lambda^a_{\delta}\int_0^\infty\zeta\e^{\beta(T-t)}\left(|x+f(t,x,a)|^{2p}-|x|^{2p}\right)\ell_{a}(a)\rd a\\
&-\sup_{\delta^{b}\in[0,k]}\lambda^b_{\delta}\int_0^\infty\zeta\e^{\beta(T-t)}\left(|x-f(t,x,b)|^{2p}-|x|^{2p}\right)\ell_{b}(b)\rd b\geq\zeta,
\end{split}
\end{equation*}
for $\beta$ sufficiently large. Now we  set
$$
v_m=\left(1-\frac{1}{m}\right) v+\frac{1}{m} v^{\zeta},\quad u_m=\left(1+\frac{1}{m}\right) u-\frac{1}{m} v^{\zeta}.
$$
Using Definition \ref{eq:def1}, one can prove that 
\begin{equation*}
\begin{split}
\min\Big\{\!\!-g(t,x)\!- \mathcal A\left( t,{\boldsymbol x},k,\partial_t\phi^{m},\partial_{s}\phi^{m},\partial_{ss}\phi^{m},\phi^m\right),&(v_m\!-\!\mathcal Mv_m)\left(t,{\boldsymbol x};k\right),\\
&(v_m\!-\!\mathcal L v_m)\left(t,{\boldsymbol x};k\right)\!\!\Big\}\geq\frac{\zeta}{m}.
\end{split}
\end{equation*}
where $\phi^{m}:=\frac{m-1}{m}\phi+\frac{1}{m}v^{\zeta}$ is the test function for $v_{m}$ and $\phi$ is the test function for $v$ and
\vspace{-0.3cm}
\begin{equation*}
\begin{split}
\min\Big\{\!\!-g(t,x)\!- \mathcal A\left( t,{\boldsymbol x},k,\partial_t\varphi^{m},\partial_{s}\varphi^{m},\partial_{ss}\varphi^{m},\varphi^m\right)      ,&(u_m\!-\!\mathcal Mu_m)\left(t,{\boldsymbol x};k\right),\\
&(u_m\!-\!\mathcal L u_m)\left(t,{\boldsymbol x};k\right)\!\!\Big\}\leq-\frac{\zeta}{m},
\end{split}
\end{equation*}
where $\varphi^{m}:=\frac{m+1}{m}\varphi-\frac{1}{m}v^{\zeta}$ is the test function for $u_{m}$ and $\varphi$ is the test function for $u$. We further note that $u$ and $v$ are polynomially bounded (see e.g. Crisafi \& Macrina (2014), Proposition 6.2, for details). Thus, we have for each $k\in\mathbb K$
$$
\lim_{{\boldsymbol x}\rightarrow\pm\infty}\left(u_{m}-v_{m}\right)(t,{\boldsymbol x};k)=\lim_{{\boldsymbol x}\rightarrow\pm\infty}\left(1+\frac{1}{m}\right)\left(u-v\right)(t,{\boldsymbol x};k)-\frac{2}{m}\zeta\e^{\beta(T-t)}(1+|x|^{2p})=-\infty,
$$
where we set $p$ larger than the bounding polynomial of $u$ and $v$. Thus the supremum is attained in a bounded  set. Since $u_{m}-v_{m}$ is upper semicontinuous,  it attains a maximum over a compact set. Next we show that, for all $m$ large, we have 
\vspace{-0.2cm}

$$
M:=\max_{t,{\boldsymbol x},k}\left(u_m(t,{\boldsymbol x};k)-v_{m}(t,{\boldsymbol x};k)\right)\leq0,
$$
\vspace{-0.2cm}
where $(\bar t,\bar{\boldsymbol x},\bar k)=\arg\max (u_m(t,{\boldsymbol x};k)-v_{m}(t,{\boldsymbol x};k))$. We define the auxiliary function $\Psi^\epsilon$ by
\vspace{-0.2cm}
$$
\Psi^{\epsilon}\left(t_1,t_2,{\boldsymbol x}_1,{\boldsymbol x}_2;k\right):=u_m\left(t_1,{\boldsymbol x}_1;k\right)-v_{m}\left(t_2,{\boldsymbol x}_2;k\right)-\frac{1}{2\epsilon}\left(|t_1-t_2|^{2}+\|{\boldsymbol x}_1-{\boldsymbol x}_2\|_2^2\right),
$$
For each $k\in\mathbb K$, $\Psi^{\epsilon}$ is upper semicontinuous and therefore it admits a maximum $M^{\epsilon,k}$ at $\big(t_{1}^{k,\epsilon},t_{2}^{k,\epsilon},{\boldsymbol x}_{1}^{k,\epsilon},{\boldsymbol x}_{2}^{k,\epsilon}\big)$. Let $M^\epsilon$ be defined by $M^\epsilon=\max_{k\in\mathbb K}M^{\epsilon,k}$, attained at the point  $(t_1^\epsilon,t_2^\epsilon,{\boldsymbol x}_1^\epsilon,{\boldsymbol x}_2^\epsilon,k^{\epsilon})\rightarrow (\bar t,\bar t,\bar{\boldsymbol x},\bar{\boldsymbol x},\bar k)$ as $\epsilon\rightarrow 0$. Furthermore we have that $M^{\epsilon}\geq M$ and $M^\epsilon\rightarrow M$ as $\epsilon\rightarrow0$. Let us assume on the contrary that $M^{\epsilon}>0$. We now go through the various cases. Let 
$$
(u_m-\mathcal Mu_m)\big(t_1^\epsilon,{\boldsymbol x}_1^\epsilon;k^\epsilon\big)\leq0.
$$
By the supersolution property of $v_m$ and by subtracting the two inequalities, we have
$$
(u_m-\mathcal Mu_m)\big(t_1^\epsilon,{\boldsymbol x}_1^\epsilon;k^\epsilon\big)-(v_m-\mathcal Mv_m)\big(t_2^\epsilon,{\boldsymbol x}_2^\epsilon;k^\epsilon\big)+\frac{\zeta}{m}\leq0
$$
We can now develop a contradiction argument since 
\begin{equation*}
\begin{split}
M&=\lim_{\epsilon\rightarrow0}u_m\big(t_1^\epsilon,{\boldsymbol x}_1^\epsilon;k^\epsilon\big)-v_m\big(t_2^\epsilon,{\boldsymbol x}_2^\epsilon;k^\epsilon\big)\\
&\leq\lim_{\epsilon\rightarrow0}\mathcal M u_m\big(t_1^\epsilon,{\boldsymbol x}_1^\epsilon;k^\epsilon\big)-\mathcal M v_m\big(t_2^\epsilon,{\boldsymbol x}_2^\epsilon;k^\epsilon\big)-\frac{\zeta}{m}\leq \lim_{\epsilon\rightarrow0}M^\epsilon-\frac{\zeta}{m}=M-\frac{\zeta}{m}.
\end{split}
\end{equation*}
The second case arises when
$
(u_m-\mathcal Lu_m)\big(t_1^\epsilon,{\boldsymbol x}_1^\epsilon;k^\epsilon\big)\leq0.
$
We  follow the same procedure to show that
\begin{equation*}
\begin{split}
M&=\lim_{\epsilon\rightarrow0}u_m\big(t_1^\epsilon,{\boldsymbol x}_1^\epsilon;k^\epsilon\big)-v_m\big(t_2^\epsilon,{\boldsymbol x}_2^\epsilon;k^\epsilon\big)\\
&\leq\lim_{\epsilon\rightarrow0}\mathcal L u_m\big(t_1^\epsilon,{\boldsymbol x}_1^\epsilon;k^\epsilon\big)-\mathcal L v_m\big(t_2^\epsilon,{\boldsymbol x}_2^\epsilon;k^\epsilon\big)-\frac{\zeta}{m}\leq \lim_{\epsilon\rightarrow0}M^\epsilon-\frac{\zeta}{m}=M-\frac{\zeta}{m}.
\end{split}
\end{equation*}
Next we consider the PIDE part and we use Definition \ref{eq:def2}. Thanks to Crandall \& Ishii's Lemma, there exist vectors such that
$$
\left(\frac{1}{\epsilon}\big(t_{1}^{\epsilon}-t_{2}^{\epsilon}\big),\frac{1}{\epsilon}\big(s_{1}^{\epsilon}-s_{2}^{\epsilon}\big),N^{\epsilon}_{1}\right)\in\bar{\mathcal P}^{-},\quad\left(\frac{1}{\epsilon}\big(t_{1}^{\epsilon}-t_{2}^{\epsilon}\big),\frac{1}{\epsilon}\big(s_{1}^{\epsilon}-s_{2}^{\epsilon}\big),N^{\epsilon}_{2}\right)\in\bar{\mathcal  P}^{+}.
$$
Recalling Definition \ref{eq:def2}, we subtract the argument of the subsolution from the argument of the supersolution and obtain
\begin{equation*}
\begin{split}
&g\big(t_{1}^{\epsilon},x_{1}^{\epsilon}\big)+\mathcal A\left(t_{1}^{\epsilon},{\boldsymbol x}_{1}^{\epsilon},k^\epsilon,\frac{1}{\epsilon}\big(t_{1}^{\epsilon}-t_{2}^{\epsilon}\big),\frac{1}{\epsilon}\big(s_{1}^{\epsilon}-s_{2}^{\epsilon}\big),N^{\epsilon}_{1},u_{m}\right)-g\big(t_{2}^{\epsilon},x_{2}^{\epsilon}\big)\\
&-\mathcal A\left(t_{2}^{\epsilon},{\boldsymbol x}_{2}^{\epsilon},k^\epsilon,\frac{1}{\epsilon}\big(t_{1}^{\epsilon}-t_{2}^{\epsilon}\big),\frac{1}{\epsilon}\big(s_{1}^{\epsilon}-s_{2}^{\epsilon}\big),N^{\epsilon}_{2},v_{m}\right)\geq\frac{\zeta}{m}
\end{split}
\end{equation*}
Since we assumed $M^{\epsilon}>0$, we choose a $\varrho>0$ such that
\begin{equation*}
\begin{split}
0&<\varrho M^{\epsilon}=\varrho\left(u_m(t_{1}^{\epsilon},x_{1}^{\epsilon};k^{\epsilon})-v_{m}(t_{2}^{\epsilon},x_{2}^{\epsilon};k^{\epsilon})-\frac{1}{2\epsilon}\left(|t^{\epsilon}_1-t^{\epsilon}_2|^{2}+\|{\boldsymbol x}^{\epsilon}_1-{\boldsymbol x}^{\epsilon}_2\|_2^2\right)\right)\\
&\leq g\big(t_{1}^{\epsilon},x_{1}^{\epsilon}\big)+\mathcal A\left(t_{1}^{\epsilon},{\boldsymbol x}_{1}^{\epsilon},k^\epsilon,\frac{1}{\epsilon}\big(t_{1}^{\epsilon}-t_{2}^{\epsilon}\big),\frac{1}{\epsilon}\big(s_{1}^{\epsilon}-s_{2}^{\epsilon}\big),N^{\epsilon}_{1},u_{m}\right)\\
&-g\big(t_{2}^{\epsilon},x_{2}^{\epsilon}\big)-\mathcal A\left(t_{2}^{\epsilon},{\boldsymbol x}_{2}^{\epsilon},k^\epsilon,\frac{1}{\epsilon}\big(t_{1}^{\epsilon}-t_{2}^{\epsilon}\big),\frac{1}{\epsilon}\big(s_{1}^{\epsilon}-s_{2}^{\epsilon}\big),N^{\epsilon}_{2},v_{m}\right)
\end{split}
\end{equation*}
We can now analyse every component in detail. First we note that, due to Assumption (vi),
$$
g\big(t_{1}^{\epsilon},x_{1}^{\epsilon}\big)-g\big(t_{2}^{\epsilon},x_{2}^{\epsilon}\big)\leq \big|g\big(t_{1}^{\epsilon},x_{1}^{\epsilon}\big)-g\big(t_{2}^{\epsilon},x_{2}^{\epsilon}\big)\big|\leq C\left(\big|t_{1}^{\epsilon}-t_{2}^{\epsilon}\big|+\big|x_{1}^{\epsilon}-x_{2}^{\epsilon}\big|\right)\rightarrow0
$$
as $\epsilon\rightarrow0$. Furthermore, by Crandall \& Ishii's Lemma and Assumption (i), we have
$$
\left(\frac{1}{2}\sigma^{2}\big(t_{1}^{\epsilon},s_{1}^{\epsilon}\big)N^{\epsilon}_{1}-\frac{1}{2}\sigma^{2}\big(t_{2}^{\epsilon},s_{2}^{\epsilon}\big)N^{\epsilon}_{2}\right)\leq \frac{C}{\epsilon}\left(\big|t_{1}^{\epsilon}-t_{2}^{\epsilon}\big|^{2}+\big|s_{1}^{\epsilon}-s_{2}^{\epsilon}\big|^{2}\right).
$$
For the integral part, we  analyse $ \mathcal B^{k^\epsilon}_{a}\big(t_{1}^{\epsilon},{\boldsymbol x}_{1}^{\epsilon},u_{m}\big)-\mathcal B^{k^\epsilon}_{a}\big(t_{2}^{\epsilon},{\boldsymbol x}_{2}^{\epsilon},v_{m}\big)$. The case involving \linebreak$\mathcal B^{k^\epsilon}_{b }\big(t_{1}^{\epsilon},{\boldsymbol x}_{1}^{\epsilon},u_{m}\big)-\mathcal B_{b }^{k^\epsilon}\big(t_{2}^{\epsilon},{\boldsymbol x}_{2}^{\epsilon},v_{m}\big)$ can be treated analogously. Given that $\sup(A)-\sup(B)\leq\sup(A-B)$, we have:
\begin{equation*}
\begin{split}
&  \mathcal B^{k^\epsilon}_{a}\big(t_{1}^{\epsilon},{\boldsymbol x}_{1}^{\epsilon},u_{m}\big)-\mathcal B^{k^\epsilon}_{a}\big(t_{2}^{\epsilon},{\boldsymbol x}_{2}^{\epsilon},v_{m}\big)\\
&\leq \sup_{\delta^{a}\in[0,k^\epsilon]}\lambda^{a}_{\delta}\int_{0}^{\infty}\Big(u_{m}\big(t_{1}^{\epsilon},x_{1}^{\epsilon}+f\big(t_{1}^{\epsilon},x_{1}^{\epsilon},a\big),y_{1}^{\epsilon}-f_{1}\big(t_{1}^{\epsilon},{\boldsymbol x}_{1}^{\epsilon},\delta^{a},a\big),s_{1}^{\epsilon};k^\epsilon\big)-u_{m}\big(t_{1}^{\epsilon},{\boldsymbol x}_{1}^{\epsilon};k^\epsilon\big) \\
&\hspace{0.75cm}-v_{m}\big(t_{2}^{\epsilon},x_{2}^{\epsilon}+f\big(t_{2}^{\epsilon},x_{2}^{\epsilon},a\big),y_{2}^{\epsilon}-f_{1}\big(t_{2}^{\epsilon},{\boldsymbol x}_{2}^{\epsilon},\delta^{a},a\big),s_{2}^{\epsilon};k^\epsilon\big)+v_{m}\big(t_{2}^{\epsilon},{\boldsymbol x}_{2}^{\epsilon};k^\epsilon\big)  \Big)\ell_{a}(a)\rd a.
\end{split}
\end{equation*}
The argument of the integral above can be rewritten as
\begin{equation*}
\begin{split}
&\Psi^{\epsilon}\big(t_{1}^{\epsilon},t_{2}^{\epsilon}, x_{1}^{\epsilon}+f\big(t_{1}^{\epsilon},x_{1}^{\epsilon},a\big),x_{2}^{\epsilon}+f\big(t_{2}^{\epsilon},x_{2}^{\epsilon},a\big), y_{1}^{\epsilon}-f_{1}\big(t_{1}^{\epsilon},{\boldsymbol x}_{1}^{\epsilon},\delta^{a},a\big),y_{2}^{\epsilon}-f_{1}\big(t_{2}^{\epsilon},{\boldsymbol x}_{2}^{\epsilon},\delta^{a},a\big),\\
&s_{1}^{\epsilon},s_{2}^{\epsilon} ;k^\epsilon\big)-\Psi^{\epsilon}\big(t_{1}^{\epsilon},t_{2}^{\epsilon}, {\boldsymbol x}_{1}^{\epsilon},{\boldsymbol x}_{2}^{\epsilon};k^\epsilon \big)-\frac{1}{2\epsilon}\Big(\big|x^{\epsilon}_1-x^{\epsilon}_2\big|^{2}+\big|y^{\epsilon}_{1}-y^{\epsilon}_{2}\big|^2\\
&-\big|x_{1}^{\epsilon}+f\big(t_{1}^{\epsilon},x_{1}^{\epsilon},a\big)  -x_{2}^{\epsilon}-f\big(t_{2}^{\epsilon},x_{2}^{\epsilon},a\big) \big|^{2}\!\!- \big| y_{1}^{\epsilon}-f_{1}\big(t_{1}^{\epsilon},{\boldsymbol x}_{1}^{\epsilon},\delta^{a}\!,a\big)- y_{2}^{\epsilon}+f_{1}\big(t_{2}^{\epsilon},{\boldsymbol x}_{2}^{\epsilon},\delta^{a}\!,a\big)  \big|^{2}\Big)\\
&\leq  -\frac{1}{2\epsilon}\Big(\big|x^{\epsilon}_1-x^{\epsilon}_2\big|^{2}+\big|y^{\epsilon}_{1}-y^{\epsilon}_{2}\big|^2-\big|x_{1}^{\epsilon}+f\big(t_{1}^{\epsilon},x_{1}^{\epsilon},a\big)  -x_{2}^{\epsilon}-f\big(t_{2}^{\epsilon},x_{2}^{\epsilon},a\big) \big|^{2}\\
&\hspace{0.4cm}- \big| y_{1}^{\epsilon}-f_{1}\big(t_{1}^{\epsilon},{\boldsymbol x}_{1}^{\epsilon},\delta^{a},a\big)- y_{2}^{\epsilon}+f_{1}\big(t_{2}^{\epsilon},{\boldsymbol x}_{2}^{\epsilon},\delta^{a},a\big)  \big|^{2}\Big),
\end{split}
\end{equation*}
where the inequality is justified by the fact that the function $\Psi^{\epsilon}$ attains its maximum at $\big(t_{1}^{\epsilon},t_{2}^{\epsilon}, {\boldsymbol x}_{1}^{\epsilon},{\boldsymbol x}_{2}^{\epsilon};k^\epsilon \big)$.
That is,
\begin{align*}
 &\mathcal B^{k^\epsilon}_{a}\big(t_{1}^{\epsilon},{\boldsymbol x}_{1}^{\epsilon},u_{m}\big)-\mathcal B^{k^\epsilon}_{a}\big(t_{2}^{\epsilon},{\boldsymbol x}_{2}^{\epsilon},v_{m}\big)\\
 &\leq\sup_{\delta^{a}\in[0,k^\epsilon]}\lambda^{a}_{\delta}\int_{0}^{\infty}\!\!\!\bigg\{-\frac{1}{2\epsilon}\Big(\big|x^{\epsilon}_1-x^{\epsilon}_2\big|^{2}+\big|y^{\epsilon}_{1}-y^{\epsilon}_{2}\big|^2
-\big|x_{1}^{\epsilon}+f\big(t_{1}^{\epsilon},x_{1}^{\epsilon},a\big)  -x_{2}^{\epsilon}-f\big(t_{2}^{\epsilon},x_{2}^{\epsilon},a\big) \big|^{2}\\
&\hspace{.75cm}- \big| y_{1}^{\epsilon}-f_{1}\big(t_{1}^{\epsilon},{\boldsymbol x}_{1}^{\epsilon},\delta^{a},a\big)- y_{2}^{\epsilon}+f_{1}\big(t_{2}^{\epsilon},{\boldsymbol x}_{2}^{\epsilon},\delta^{a},a\big)  \big|^{2}\Big)\bigg\}\ell_{a}(a)\ \rd a.
\end{align*}
Since the right-hand-side of the above equality tends to zero as $\epsilon\rightarrow0$, we have that $\lim_{\epsilon\rightarrow0} \mathcal B^{k^\epsilon}_{a}\big(t_{1}^{\epsilon},{\boldsymbol x}_{1}^{\epsilon},u_{m}\big)-\mathcal B^{k^\epsilon}_{a}\big(t_{2}^{\epsilon},{\boldsymbol x}_{2}^{\epsilon},v_{m}\big)\leq0$. Finally, we have
$$
\mathcal Q\Big(u_{m}\big(t_{1}^{\epsilon},{\boldsymbol x}_{1}^{\epsilon};k^\epsilon\big)-v_{m}\big(t_{2}^{\epsilon},{\boldsymbol x}_{2}^{\epsilon};k^\epsilon\big) \Big)\leq0
$$
since the maximum is attained at $k^\epsilon$. Thus, by letting $\epsilon\rightarrow0$, we get $\varrho M\leq0$, which is a contradiction since $\varrho>0$. Therefore $M\leq0$. Furthermore, since we have proved that $u^{*}\leq v_{*}$, the value function is  continuous as it is both upper and lower semicontinuous.\hfill$\square$

\end{document}